\documentclass[12pt]{article}
\usepackage[utf8]{inputenc}
\usepackage{amsmath}
\usepackage{multicol}
\usepackage{authblk}

\usepackage[backend=biber,style=phys]{biblatex}
\usepackage{graphicx}
\usepackage{subcaption}
\usepackage{multirow}

\usepackage[left=2cm, right=2cm, top=2cm, bottom=2cm]{geometry}

\title{Electronic Structure of First and Second Row Atoms under Harmonic Confinement}

\author[1,2]{A. Robles-Navarro}
\author[1,2]{P. Fuentealba}
\author[1,2]{F. Mu\~noz}
\author[1,2]{C. C\'ardenas\thanks{Corresponding Author: cardena@uchile.cl}}
\affil[1]{Departamento de F\'isica, Facultad de Ciencias, Universidad de Chile, Casilla 653, Santiago, Chile}
\affil[2]{Centro para el Desarrollo de la Nanociencia y la Nanotecnolog\'ia (CEDENNA), Avda. Ecuador 3493, Santiago 9170124, Chile}

\begin{document}
\maketitle

\begin{abstract}
Atoms under confinement induced, for instance, by high external pressure, undergo a series of processes and modification of its electronic structure. Good examples are the spontaneous ionization, stabilization of excited-state configurations that result in a level-crossing with the ground state of the free atom, contraction of atomic shells, and different regimes of electron correlation. In this work, we do a systematic study of the effects of confinement with harmonic potential on the electronic structure of atoms from H to Ne. Dynamic electron correlation is taken into account by performing CCSD calculations. When necessary, static correlation is treated with CASSCF. Because the strength of harmonic confinement cannot be translated into pressure, we envisioned a "calibration" method for establishing an adequate volume to transform confinement into pressure. 
 We focused on the effect of confinement on the following properties: i) changes of electron distribution and electron localization within the $K$ and $L$ atomic shells, ii) confinement-induced ionization pressure, iii) level crossing of electronic states, and iv) the behavior of the electron correlation energy. We found that contraction of valence and core shells are not negligible and that the use of standard pseudopotentials might be not adequate to study solids under extreme pressures. The critical pressure at which and atom ionizes follows a periodic trend, and it ranges from $28$ GPa for Li to $10.8$ TPa for Ne. In Li and a Be, pressure induces mixing of the ground state configuration with excited states. At high pressure, the ground state of Li and Be becomes a doublet and a triplet with configurations $1s^2 2p$ and $1s^2 2s 2p$ respectively. The potential consequences of these changes of configuration on the chemistry of Be are discussed. Finally, the changes in the amount of electron correlation are characterized and analyzed in terms of the RPA approximation. For atoms with fewer electrons in the valence shell correlation increases, but for atoms with more electron, the increasing of kinetic energy dominates over electron correlation. 

\end{abstract}
\newpage

\section{Introduction}

A matter of research interesting to physics and chemistry is the study of electronic systems (atoms, molecules, clusters, and solids) under extreme conditions such as high pressures or confinement. Electronic systems under confinement are not only crucial for simulation of the effects of high-pressure on electronic properties, but they also are interesting by itself in the study of quantum dots or encapsulated atoms or molecules. Examples are atoms or molecules encapsulated in cages like fullerenes or zeolites. The confinement sometimes produces essential changes in the electronic structure of the system. It affects its bonding pattern, its possible catalytic properties and in the solid state can dramatically change the stable crystallographic phases. One interesting example is the high-pressure electrides, for which a unified theoretical model have been presented\cite{Hoffmann2014}.  

The physical and chemical properties of systems under high pressure (strong confinement) can be quite different from their free partners and elude intuition. For instance, at enough high pressure, atoms change their valence state, implying, for instance, a change in the coordination number and the appearance of new phases of solids under pressure. This happens because in atoms, there is a crossing between different energy levels and orbitals that are unoccupied in the free atom could be filled in the compressed one. This type of transitions were experimentally detected\cite{Takemura1982,Takemura1985,Tups1982}. An excellent review of those effects in solids has the striking title of “The Chemical Imagination at Work in Very Tight Places”\cite{Grochala2007}.  Another known phenomenon occurs when the $PV$ term ($P$ is the pressure and $V$ the volume) in the equation-of-state works against the binding of the electron and confinement-induced ionization occurs. Connerade wrote an interesting review of the topic\cite{Dolmatov2004}. There are a variety of computational and experimental studies on the effect of high pressure on phase stability of solids\cite{recio2016book}.

Electronic confinement can be explicitly simulated by including the agent that exerts the reduction of the available volume, such as placing an atom inside a molecular cage or simulating high pressure by changing the length of the unitary cell in a calculation with periodic boundary conditions. Nevertheless, if the confinement is strong or there is a substantial overlap of the wavefunctions of the system under confinement and its surroundings, then it makes it difficult to differentiate between both of them. Hence, modeling the confining agent with an external potential has the advantage that the intrinsic response of the system to confinement can be always characterized. Two types of potentials are most used: \textit{i)} an infinite potential on the surface of a cavity which imposes Dirichlet boundary conditions to the wavefunction (hard walls), and \textit{ii)} soft (penetrable) potentials which allows the wavefunction to penetrate in classically forbidden regions. A prototypical example of the last type of potentials is the harmonic oscillator, and it is the one used in this work. 

In a series of papers, Diercksen et al. studied the electronic states and their density for low-lying states of atoms and diatomic molecules confined by harmonic oscillator potentials\cite{Diercksen2001,Diercksen2005,Diercksen2005book,Diercksen2008,Sako2005}. They looked for different degrees of confinement varying the frequency of the oscillator. Perhaps the most important point of their works was the use of very accurate wave functions. For instance, for He atom they employed a full configuration interaction, CI, with a pervasive basis set, and for Li atom an extensive multireference configuration interaction wavefunction. Very interestingly, they showed that  Gaussian basis set could be extremely accurate to expand the atomic orbitals in the presence of harmonic confinement. The basis set, however, has to be balanced in the sense that it should provide a good representation of both, the bound states of the Coulomb potential and the bound state of the harmonic oscillator. For instance, for Li they used a basis set as large as [13s7p5d]+[1s1p1d1f1g1h]. The results they obtained are significant as a benchmark for the use of simpler models. They also show that Gaussian basis sets are suitable for a quasi-two dimensional attractive Gaussian quantum dot.

Cioslowski et al.\cite{cioslowski2014,Cioslowski2015,Matito2010,Matito2015} have studied the effects of harmonic confinement in the so called harmonium atoms. One striking finding of the authors is the emergence of  Wigner molecules in three-dimensional Coulombic systems, which takes place over several orders of magnitude of confinement\cite{cioslowski2017,cioslowski2006}. 

Years ago, Chattaraj et al. began to study the effects of confinement on the chemical reactivity\cite{Chattaraj2003}. One of their first work was on the chemical reactivity of atoms confined in a spherical box. They calculated the variations of some indices derived in the density functional theory of chemical reactivity and the variation of the atomic dipolar polarizability. They did numerical calculations with Dirichlet boundary conditions and found that in general, the dipolar polarizability decreases with the increase of the confinement, a very consistent result. Later on, they used a variety of theoretical methodologies to show that confinement has a significant effect on many  classical chemical reactions. For example, they used a relative big molecular host to confine some model Diels-Alder reactions and find the catalytic changes due to the confinement\cite{Chattaraj2017catal,Chattaraj2018catal,Chattaraj2018catal2}. The theoretical models they used range from numerical Hartree-Fock to more sophisticated quantum fluid dynamics time-dependent density functional methods developed by his group. Very recently, they wrote a feature article summarizing their results\cite{Chattaraj2019rev}.

Other kind of works have been done by Garza et al.\cite{Garza2005,Bautista2017,garza2016,AquinoGarza2006,Bautista2015}. They also began with the study of confined atoms in a rigid spherical wall. However, they interpreted the results in terms of pressure by using the thermodynamic relation $P=-\frac{\partial E}{\partial V}$ and they also derived a numerical equation to calculate the Gibbs free energy. They obtained variations of electronic properties under pressure changes\cite{guerra2009}. Later on, they extended the results to the pressure changes with soft spherical walls. They also constructed special basis sets to be used with this potential. This is a critical point many times overlooked (exceptions are the cited works by Diercksen et al.). Note that commonly-used Gaussian basis sets for free atoms do not take into account that a wall/potential changes the way the wavefunction decays. Furthermore, the contraction scheme of the core in atomic-free basis set does not give enough flexibility to the wavefunction to capture the reorganization of the inner region of atoms under strong confinement. 

This work studies the effects of the confinement/pressure on the electronic structure of atoms from Hydrogen to Neon. Electrons are confined with an isotropic harmonic potential. Variation of the total energy and the electron density under confinement are discussed and a scheme to translate confinement strength into pressure is introduced. The electron's distribution are also analyzed with the aid of the  Electron Localization Function (ELF), which allows us to discuss the changes in the electron-shells due to the confinement. It is found that the required pressure to ionize the atom follows a  periodic behavior. In some atoms occur an energy level crossing before ionization.  It is also found that the impact of confinement on the electronic correlation depends on the number of electrons of the atoms and its behavior is semi-quantitatively explained in terms of the RPA model of correlation energy at the high-density limit. To compare the importance of the degree of approximation in the solution of the electronic Hamiltonian, 
Hartree-Fock (HF), Kohn-Sham with the PBE exchange-correlation functional (DFT), and Coupled Cluster with single and double excitations (CCSD) calculations have been implemented.  The change of configuration of Li and Be with confinement is characterized by CASSCF calculations, and it is  discussed in detail.

\section{Computational Methods}

The Born-Oppenheimer Hamiltonian of atoms under isotropic harmonic confinement is, in atomic units,
\begin{equation}
\hat{H} = -\frac{1}{2} \sum \nabla^2 _i - \sum \frac{Z}{r_i} + \frac{1}{2} \sum \frac{1}{r_{ij}} + \frac{1}{2} \sum \omega^2 r^2 _i,
\end{equation}
where the first term is the electron kinetic energy, followed by the nuclei-electron attraction and the electron-electron repulsion. The last term is the harmonic potential that confines electrons and it is centered in the nucleus position. The parameter $\omega$ controls the degree of confinement. The basis set for expanding atomic orbitals should include both, functions for the bound states of the Coulomb potential and functions for the states of harmonic oscillator. For the part of harmonic oscillator we chose Gaussian functions with suitable exponential. Diercksen et al.\cite{Sako_2003} found that  the optimal exponents follow the approximated series $\omega$, $\omega/2$, $\omega/4\dots\omega/2n$. We have used the first four exponents of the series and included basis set with angular momentum $l=0$, $1$ and $2$. For the Coulombic part of the potential  we used a decontracted 6-311G(d,p) basis set. To have an idea of the size of  basis set of this scheme, in the case of Fluorine there are 67 basis functions. We expect the effect of the basis set to be more important in atoms with more electrons and open shell. Hence, for Fluorine we check the convergence of the CCSD energy with the type of angular momentum basis set of the harmonic potential by increasing the angular momentum up to $l=5$ (see Figure S2 in supplementary material).  It is observed that for $\omega<0.4$ the energy is converged to $3\times10^{-4}\ E_h$. For the confinement strength for which ionization occurs, the error is $8\times10^{-3}\ E_h$.  
All the necessary integrals and diagonalization of Hamiltonians were done using tools of the quantum chemistry code Gaussian09\cite{g09}. Electron density and ELF analysis was done with HORTON\cite{horton} and ChemTools\cite{HeidarZadeh2016} programs. The calculation of the ELF in correlated wavefunctions (CCSD) was done in term of the natural orbitals that diagonalize the variational (relaxed) density matrix, following the procedure of Matito et al.\cite{Matito-NO-ELF} 


Different levels of theory were used. Hartree-Fock, Coupled Cluster with all (core included) single and double excitations, CCSD, and Kohn-Sham calculations with the PBE exchange-correlation functional\cite{Perdew1996}. Further, for Li and Be atoms, a CASSCF calculations with 4 active orbitals and 1 and 2 electrons, respectively, were done around the $\omega$ value at which crossing of states occurs. 

Confinement with penetrable walls, contrary to hard walls, comes with the difficulty of defining the pressure associated with a given strength of confinement. However, the pressure is a measurable quantity, while the strength of confinement ($\omega$) is a parameter of a model. 
To express confinement in terms of pressure instead of $\omega$, one needs a measure of the volume. When confinement is done with hard walls the volume is well defined. Though, in the case of penetrable walls, a criterion has to be chosen to select the volume of the confined atom. In  calculations with harmonic confinement we observed that if one defines as a volume that of a sphere enclosing most of the electron density, then the pressure nicely follows the thermodynamic definition, namely,  a linear fit between  the derivative of the energy with respect to the inverse of the volume and the square of it: 
\begin{equation} \label{eq:eos}
    P = \frac{1}{V^2}\left(\frac{dE}{d(V^{-1})}\right).
\end{equation}

The quality of the fit is almost independent ($R^2 > 0.999547$) of the volume provided that the sphere encloses more than $90\%$ of the electrons. To have a non-arbitrary scale of pressure, we tuned the radius of the sphere such that the pressure of ionization of H in the harmonic potential equals that of the H in hard walls. This pressure,  $620$ GPa,  has been accurately determined by Aquino\cite{Aquino1995} and Rubinstein et al.\cite{Rubinstein1979}. We found that the best sphere contains  $96.648\%$ of the electrons. Therefore, in all results we present for other atoms, the pressure is computed by fitting  Eq. \ref{eq:eos} with the volume defined as above. However, all tables are also available  in terms of $\omega$ in the supplementary material.

Additionally, the electron localization function is also calculated. This function has been extensively studied and various reviews of it exist\cite{Savin1991,Savin1997,Savin2005,Silvi2005,Fuentealba2007,Burdett1998}. Here we use the interpretation of the ELF due to Savin, which says that the ELF is a measure of the excess of kinetic energy due to the Pauli principle. As a consequence, values of the ELF close to one represents the regions of the space where is more probable to find localized electrons.

\section{Results and Discussion}

\begin{figure}[ht]
    \centering
    \includegraphics{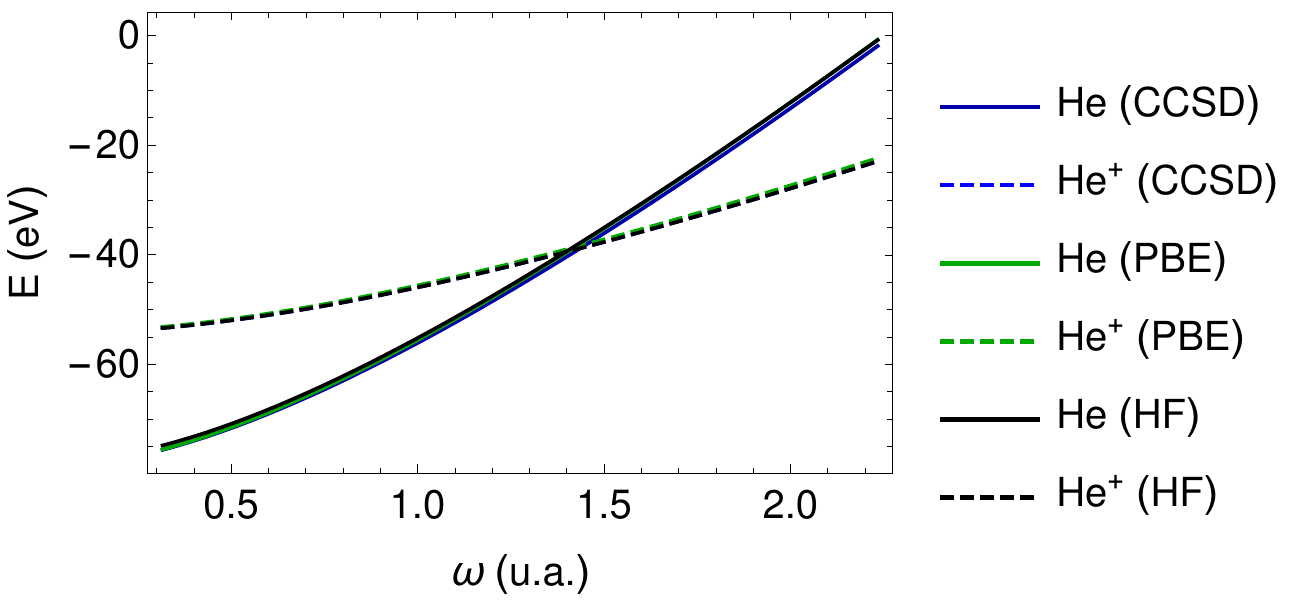}
    \caption{Energy variation of the He atom with increasing confinement.}
    \label{fig:Ew_He}
\end{figure}

Because discussing the effect of confinement on the electronic structure of each atom could be lengthy and unnecessary, in this section we discuss only some representative results for different atoms while the full results for H to Ne can be found in the supplementary material. Let us start with the variation of the energy of He atom as the confinement increases (Figure \ref{fig:Ew_He}). 

It is first to notice that the way  the energy varies with $\omega$ is quite independent of the method of calculation, from HF to CCSD calculations, indicating that in He the effects of electron-electron correlation remain small when compared with the total energy. However,  the correlation energy at the strongest confinement could be the double than the correlation energy in free atoms. This is further discussed at the end of this section. For atoms different than He, the $E$ vs. $\omega$ curve follow the same trend as far as there is not a level-crossing in the range of $\omega$.  Note that the energy of both, the neutral atom and its cation increases with $\omega$. Nevertheless, the energy of cation increases less sharply that the neutral. Therefore, there is a  value of $\omega$  at which both energies equal. At this point one says that ionization (some author call it autoionization\cite{Dolmatov2004}) occurs because beyond this point the energy of the cation He$^+$ is less than the energy of the neutral. It can be seen from  Figure \ref{fig:Ew_He} that this happens at $\omega=1.44$ a.u. which corresponds to a pressure of $5535$ GPa. As expected a value much higher than the one for the hydrogen atom ($620$ GPa)

\begin{figure}[ht]
    \centering
    \includegraphics[scale=0.7]{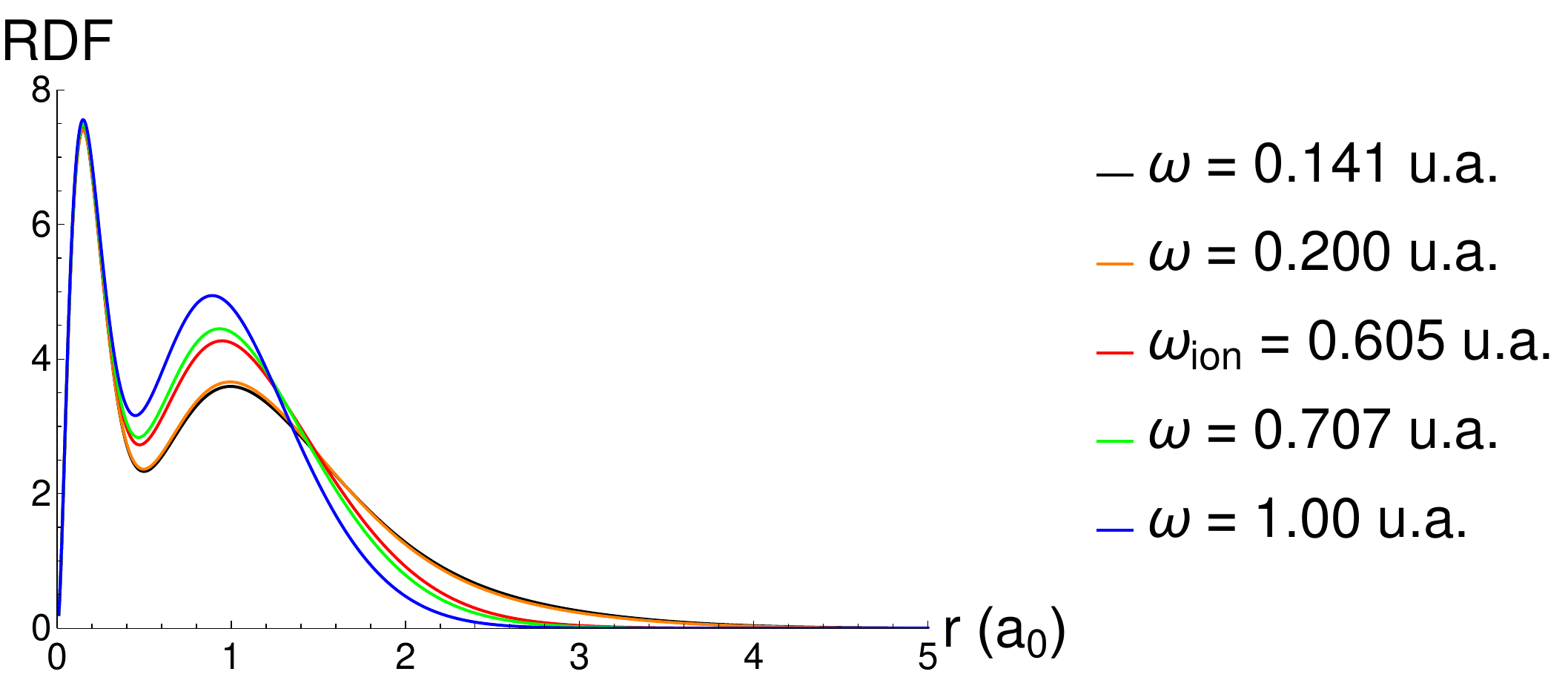}
    \caption{Radial distribution function of the Nitrogen atom at different confinements.}
    \label{fig:RDF_N}
\end{figure}

To look at the electron density variations, in Figure \ref{fig:RDF_N}, the radial distribution of the electron density of the Nitrogen atom at different values of the confinement parameter is plotted. 
The density decays faster as the confinement increases, and the maximum of the outer shell ($L$)  gets more compact and, therefore, denser. The density at the minimum that separates the core shell ($K$) from the valence shell ($L$)  also increases with the confinement, making the population of both shells more correlated. Also, the size of the $K$ shell slightly decreases with confinement. It is to recall that similar results have been obtained for all atoms of the first and second periods.

\begin{figure}[ht]
    \centering
    \includegraphics[scale=0.7]{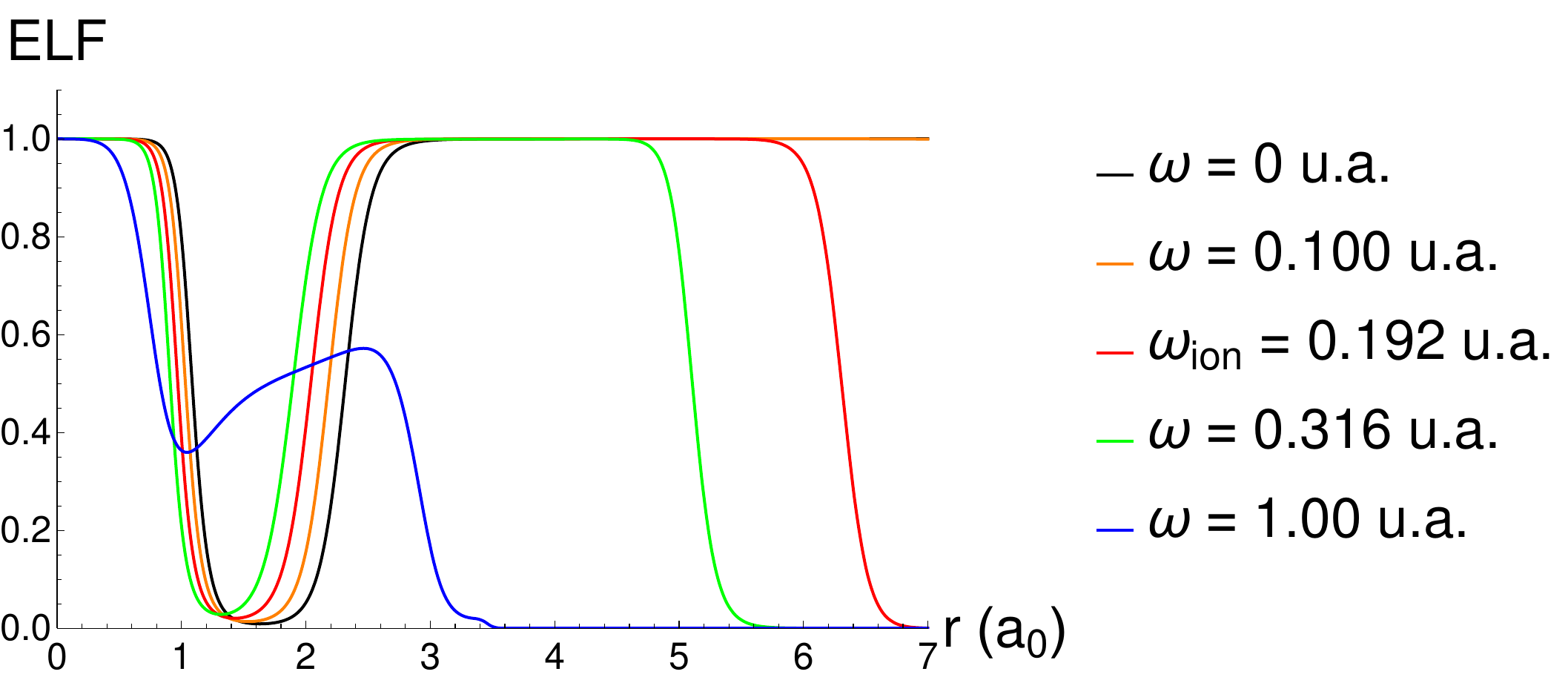}
    \caption{Electron localization function of the Lithium atom at different strength of confinements. $\omega_{\text{ion}}$ corresponds to the critical confinement strength at which the atom ionizes.}
    \label{fig:ELF_Li_CCSD}
\end{figure}

Variations of the electron distribution and structure of the shells are more easily captured with the ELF, which is shown for the Lithium in Figure \ref{fig:ELF_Li_CCSD}. Again, the ELF shows that the core of the atom compacts upon confinement and that its size, measured by the position of  minimum between $K$ and $L$ shells, decreases almost a Bohr for the range of confinement of Figure  \ref{fig:ELF_Li_CCSD}.

Note that the ELF of free Lithium, and other alkaline metal atoms, is a particular case as it does not go asymptotically to zero as the distance to the nuclei goes to infinity. This behavior is because the perfectly symmetric one-electron outer shell. However, the ELF of Lithium under enough confinement does go to zero with the distance because of the fast decreasing of the density.  Interestingly, the pattern of the ELF is quite different at strong confinement ($\omega=1$). Electron in the $L$ shell is as delocalized as an electron in the non-interacting homogeneous gas (ELF=$0.5$). Delocalization also notoriously increases in the core shell, and its size decreases from almost  2$a_0$ in the free atom to 1$a_0$ at $\omega=1$. This change in electron localization at strong confinement is associated with a change in the electron configuration of the atom, which passes from $1s^22s$ to $1s^22p$. A CASSCF calculation with 1 electron and 4 active orbitals reveals that around $\omega=0.806$, both states become degenerate ($\Delta E<0.3$ eV), and the atom lives in a mixed state of valence. The crossing between these two states was also predicted for the Lithium atom in the hard sphere model\cite{Patil2004}.
 
\begin{figure}[ht]
    \centering
    \includegraphics[scale=0.7]{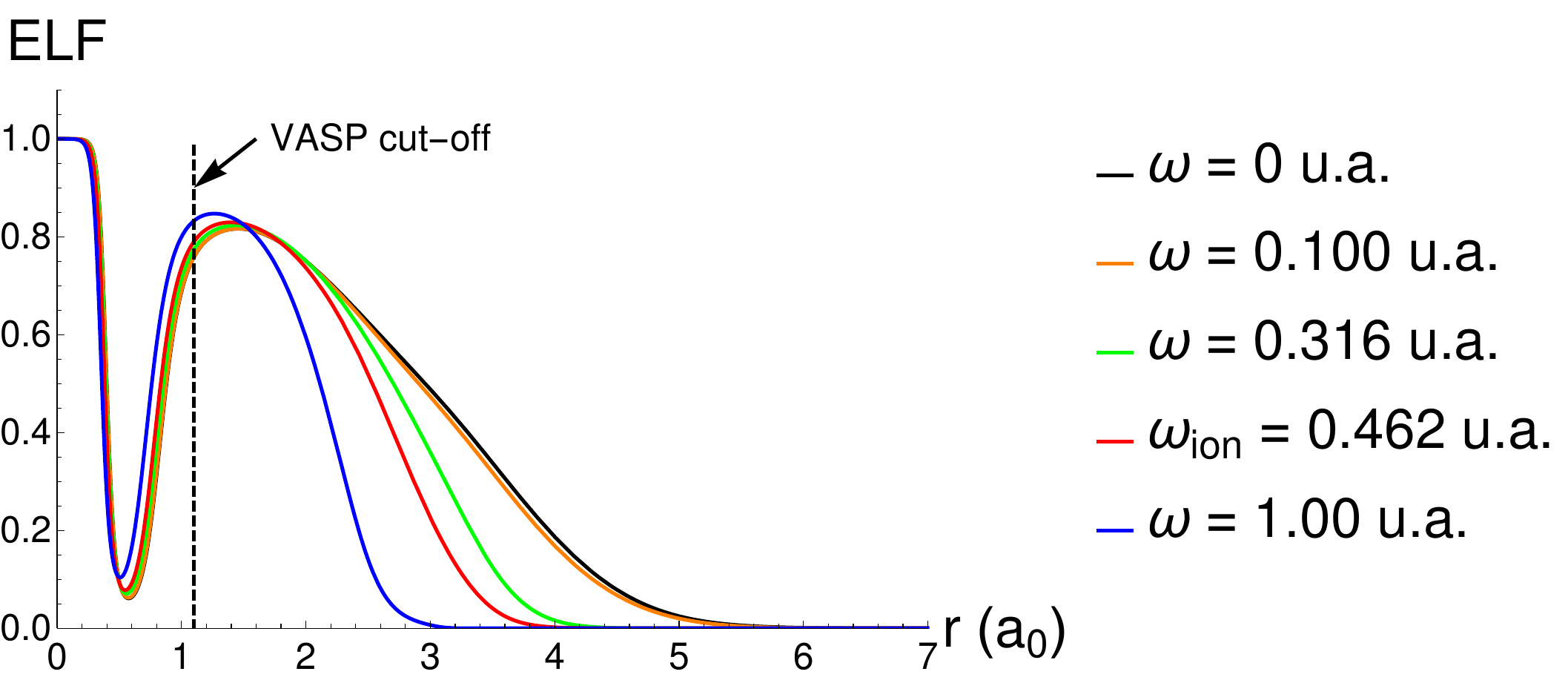}
    \caption{Electron localization function of the Carbon atom at different confinements.}
    \label{fig:ELF_C_CCSD}
\end{figure}

The ELF of Carbon under confinement is shown in Figure \ref{fig:ELF_C_CCSD}. Here, it is easier to see how the shells contract with confinement. The boundary of the $L$ shell decays sharply from $\approx5.6$ $a_0$ for the free atom to $\approx3.2$ $a_0$ for $\omega=1$. Also, as the outer shell gets closer to the nucleus, it also squeezes the inner shell. The position of the maximum of the outer shell is quite affected by confinement as it changes from  $\approx1.6$ $a_0$ for the free atom to $\approx1.2\,a_0$ for $\omega=1$. These changes in the electron distribution of the outer shell may have been being overlooked when standard pseudo-potentials are used at very high pressures. For instance, the dashed vertical line in Figure \ref{fig:ELF_C_CCSD} corresponds to the  cutoff radius ($1.1\ a_0$) of the PAW pseudopotetial of C with the smallest core in the  VASP program\cite{Kresse1999}. Hence, for pressures greater than $300$ GPa ($\omega=0.1$), the changes of electronic structure in the sphere defined by the cutoff radius are not negligible. It means that under enough pressure, the core-valence separation in pseudo-potentials needs to be tailored.\cite{benedict2014}

\begin{figure}[ht]
    \centering
    \includegraphics[scale=0.75]{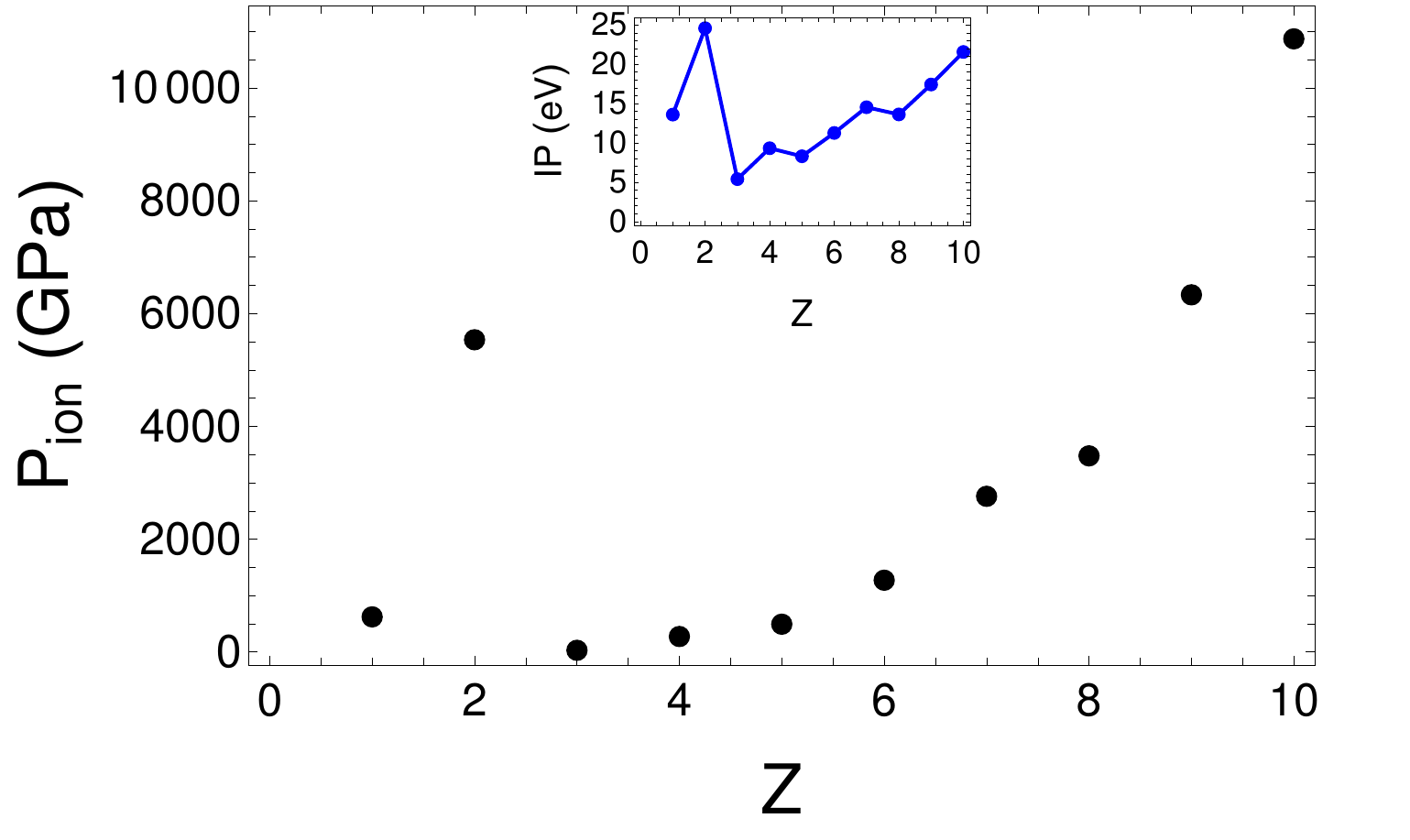}
    \caption{Ionization pressure of the atoms from H to Ne. As an inset, the ionization potential of the first and second row atoms.}
    \label{fig:P_vs_Z}
\end{figure}

Figure \ref{fig:P_vs_Z} shows the necessary pressure to ionize the atoms from H to Ne calculated with the CCSD method. Notice the perfect periodic variation of the ionization pressure, which is largest for the noble gas atoms (He and Ne). Not surprisingly,  Lithium has the smallest ionization pressure ($27$ GPa). It is so small compared to other atoms that in the scale of the plot it seems to be zero. It is interesting to note that periodic behavior of the ionization pressure it is so similar to other properties, such as the ionization potential,  that it even shows the small kink of the Nitrogen atom in the second period.\cite{cardenas2016}

Because ionization pressure depends on the total energies of the atom and its cation, it does not depend strongly on the method of calculation, HF, CCSD or PBE. However, there is a trend of PBE to overestimate the pressure slightly, while HF does the opposite. This can be understood because GGA functionals suffer from a delocalization error opposite to HF.\cite{wyang2008} That is, DFT tends to delocalize electrons while HF tends to localize them. Delocalize electrons have high kinetic energy and, therefore, high pressure. This suggest that a LDA and a HF calculation could be used to set "errors bars" to computational phase-transition pressures in solids. Table 1  summarizes the ionization pressure for all atoms. When interpreting the values of that table,  it is important to keep in mind that those values have uncertainty associated with the statistical error of the fitting of Eq. \ref{eq:eos}  and that $1$ GPa is only $3.4\times10^{-5}$ a.u.. Interestingly, our ionization pressures agree with the prediction of Hoffmann et al. \cite{Hoffmann2014} that no element of the second period ionizes for pressures below than $500$ GPa but Lithium, which does below  $80$ GPa. We also estimate the ionization pressure of Li from recent HF calculations with hard walls done by Rodriguez et al. \cite{Bautista2017}. Despite the restricted data, we found that the ionization pressure ($30$ GPa) agrees very well with our HF value ($27$ GPa). This nice agreement strongly support our scheme to translate values of $\omega$ into pressure. We agree on the ionization pressures despite the different nature of the confinements.

Data equivalent to Table 1, but written in terms of $\omega$, is reported in Table S1 in the supplementary material. This information is more useful for comparison with future calculations. Table S1  also shows the ionization $\omega$ for the Koopmans' approximation within HF. If CCSD is taken as a reference value, the mean absolute percentage error the of the ionization $\omega$ is $2.1\%$, $2.5\%$, and $5.5\%$ for HF, PBE and Koopmans' approximation respectively. Special mention deserves the error of Koopmans' approximation because it is commonly used to estimate ionization and crossing of different atomic states.  Although Koopmans' approximation is qualitatively correct, quantitatively its error is not negligible.

\begin{table}[ht]
\centering
\label{tb:pressure-ion}
\caption{Ionization pressure of the first and second row atoms.}
\begin{tabular}{ccccc}
\hline
\multirow{2}{*}{Atom} & \multicolumn{3}{c}{$P_\text{ion}\text{(GPa)}$} \\ \cline{2-4} 
                      & HF         & CCSD      & PBE       \\ \cline{1-4} 
H                     & $621  $    & ---       & ---       \\
He                    & $5262$     & $5535$    & $5541$    \\
Li                    & $27.0$     & $27.8$    & $31.8$   \\
Be                    & $216$      & $273$     & $262$   \\
B                     & $511$      & $491$     & $540$   \\
C                     & $1279$     & $1271$    & $1308$    \\
N                     & $2757$     & $2759$    & $2756$    \\
O                     & $3386$     & $3476$    & $3640$    \\
F                     & $6202$     & $6332$    & $6437$    \\
Ne                    & $10656$    & $10871$   & $10864$   \\ \hline
\end{tabular}
\end{table}

\begin{figure}[ht!]
    \centering
    \includegraphics[scale=0.4]{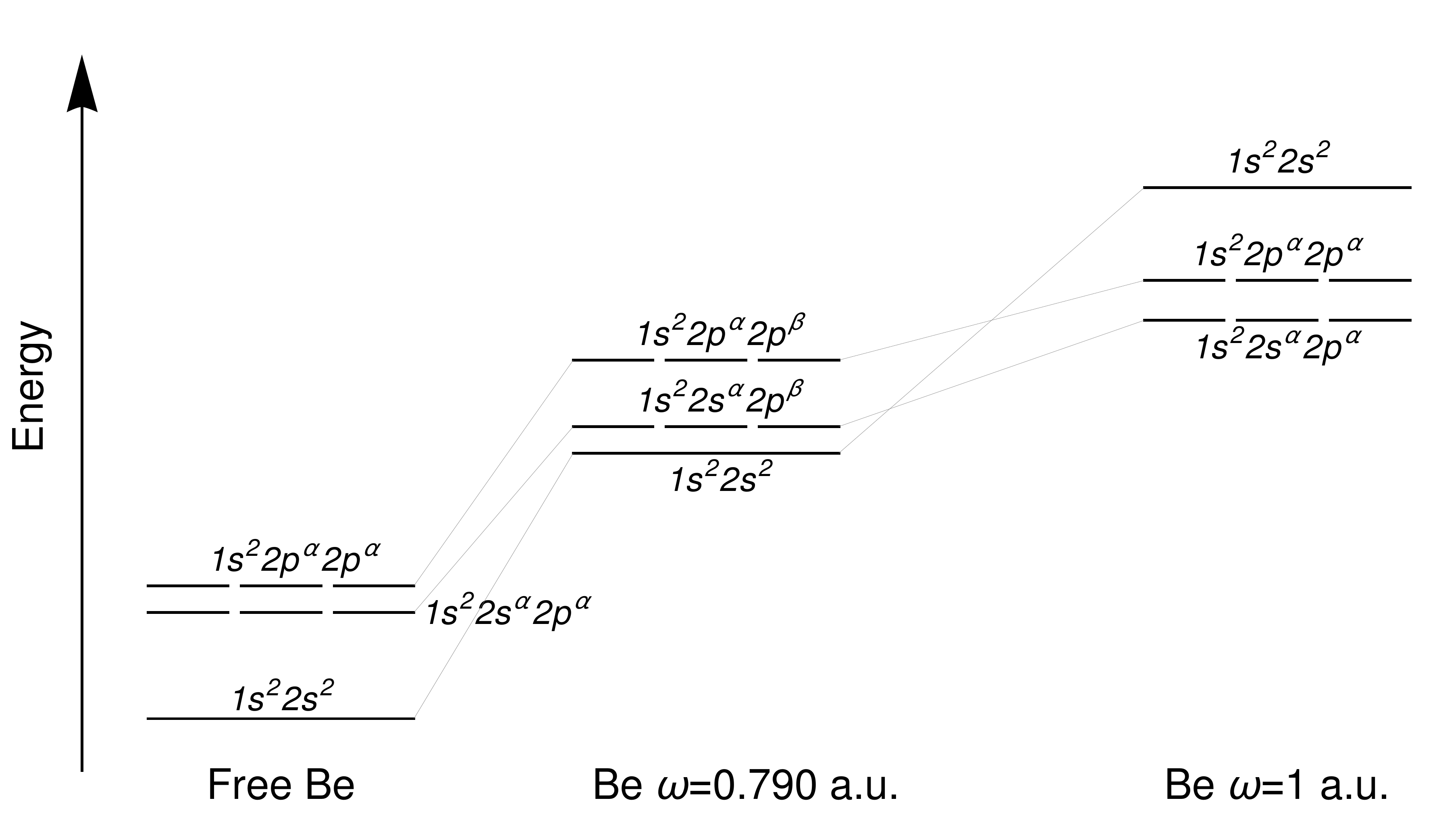}
    \caption{Lowest configurations of the Be atom at different confinements.}
    \label{fig:CAS_Be}
\end{figure}

Beside ionization, confinement can also induce a level crossing. That is, the electronic configuration of the ground state of the confined atoms can be different from the free atom. In an orbital picture, such as in  HF and DFT, this manifest in the change of order of the atomic orbitals  energies. For instance, for Be, there is critical confinement for which the orbitals $2s$ and $2p$ becomes degenerate. This level crossing has already been reported with HF and DFT methods.\cite{Sanu2018,Garza2005} However, in a degenerate state, no single determinant can be a good approximation to the quantum state because ``static'' correlation becomes dominant. Hence,  we resort to CASSCF calculations with 2 electrons and 4 active orbitals (2s and 2p) to find the best value of $\omega$ for which the crossing occurs and the electronic configuration of the ground state and its (pseudo-)degenerate states. The crossing occurs around  $\omega=0.790$, (See  In Fig. \ref{fig:CAS_Be}). In the free atom, the first two excited state, two triplets with configuration $1s^22s2p$ and $1s^22p^2$, are well separated ($2.42$ and $7.14$ eV) from the singlet ground state $1s^22s^2$. Contrary, at $\omega=0.790$ the singlet $1s^22s^2$ is degenerate with the triplet  $1s^22s2p$ ($\Delta E<0.3$ eV), while the other triplet lies only $3.22$ eV above the degenerate states. When the confinement increases beyond  $\omega=0.790$, the ground state of Be becomes the triplet $1s^22s2p$, as it can be seen for $\omega=1.0$ in  Fig. \ref{fig:CAS_Be}). The ground becomes a  triplet as a result of the Hund's rule, which operates independently of the strength of the confinement. Sarsa et al.\cite{Sarsa2018} have recently addressed this matter. If the confinement keeps increasing beyond $\omega=1.1$, the low lying excited state  $1s^22p^2$ becomes the ground state, which agrees with HF calculations recently reported by Sa\~nu-Ginarte et al.\cite{Sanu2018} for hard spherical walls. Note that capturing correlation energy of free Be is a classic, challenging situation because both static and dynamic correlations are important. Therefore, that HF, CCSD and CASSCF  calculations predict the same ground state in the limit of strong confinement (with soft and hard walls) reinforces the idea that in that limit the increasing of kinetic energy prevails over correlation energy.

The chemical importance of level crossing is not to diminish. At a pressure equivalent to the confinement of the crossing, the chemistry of Beryllium atom would be completely different. It will behave not longer as an earth alkaline metal atom. It would be a very reactive species able of $sp$ hybridization that would resemble Boron. Of course, at high-pressure one does not expect to have a gaseous phase. However, the new configuration would dictate  new solid-state phases or exotic molecules. 

\subsection{Correlation energy}

\begin{figure}[ht]
    \centering
    \includegraphics[scale=0.8]{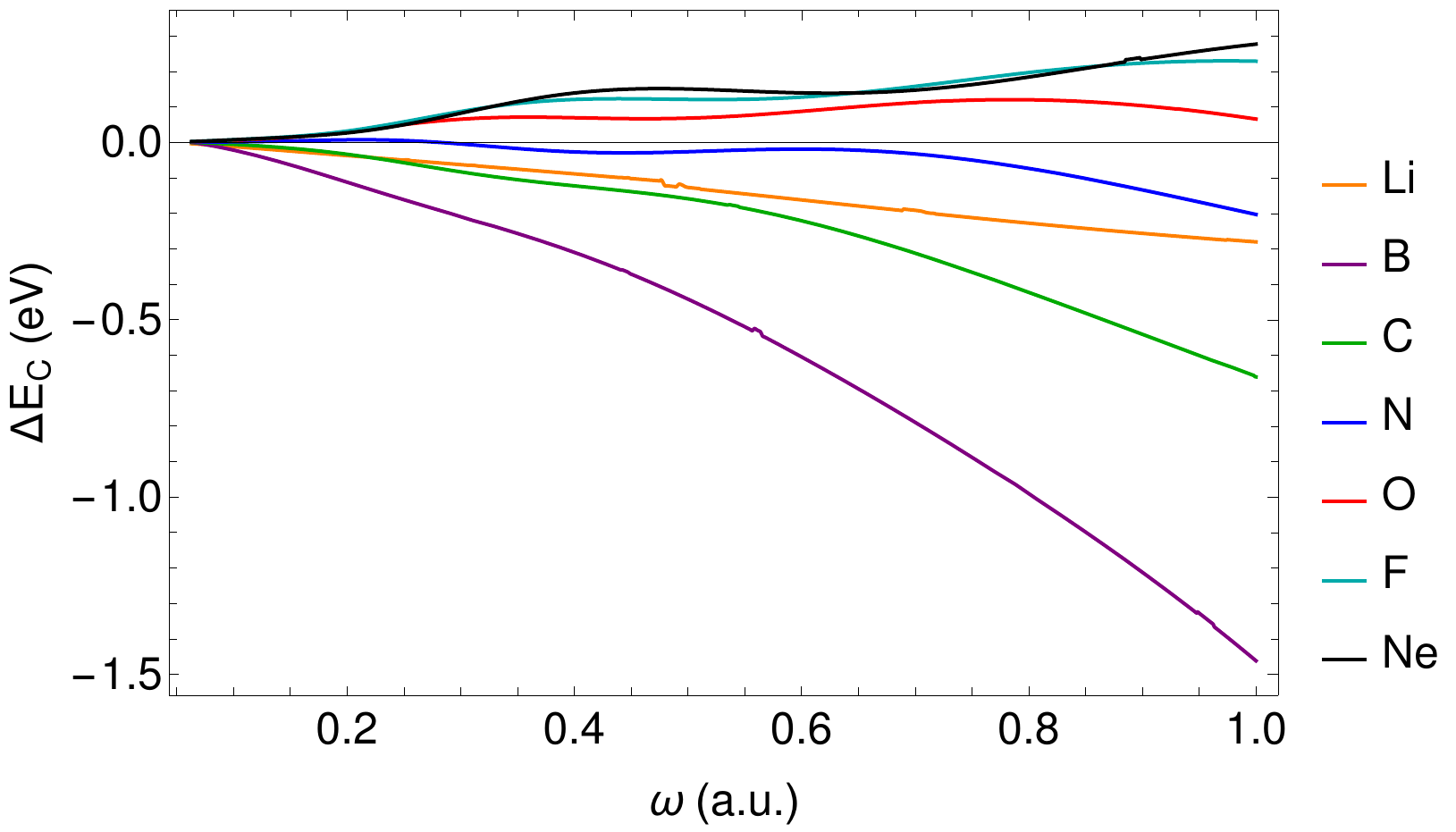}
    \caption{Change in correlation energy with confinement for second row atoms.}
    \label{fig:decorr}
\end{figure}

Most calculations of confined atoms resort on  HF or DFT methods. Therefore, there is little numerical evidence on the effect of the confinement on the ab initio correlation energy ($E_{\text{exact}}-E_{\text{HF}}<0$). CCSD is, within the basis set, and exact solution for He and it recover most of the correlation energy for atoms of the second period. An exception is Be because in this atom static correlation is crucial and CCSD includes mostly dynamic correlation. Hence, excluding Be, one observes that confinement can either increase or decrease correlation energy. For Ne, F, and O the correlation energy  becomes less negative (increases) with confinement, while it becomes more negative (decreases) for Li, B, C and N.  The change, with respect to the free atom, of  the correlation energy as a function of $\omega$ is plotted in Figure \ref{fig:decorr}. The correlation becomes significant for atoms with fewer electrons. For instance, for the strongest confinement $(\omega=1.0)$, the correlation energy of Boron decreases $1.5$ eV ($0.055\ E_h$), which corresponds to a change of 152\% of the correlation energy with respect to the free atom. On the contrary, in Ne and F, the reduction of correlation translates in an increment of 96\% of the correlation energy. In other atoms, the perceptual change is $-79\%$ for Li, $-120\%$ for C, $-105\%$ for N, and $99\%$ for O.  


The dependence of the correlation energy on the confinement-strength and number of electrons reveals that for a given confinement-strength, in the atoms with more electrons, the increment of the kinetic energy with the density becomes dominant over the correlation energy. This behavior can be understood in a semi-quantitative way with the RPA model of the kinetic and correlation energy of the homogeneous electron gas (HEG) in the high-density limit\cite{mahan-many}. In the HEG the kinetic energy is given by

\begin{equation}
\epsilon_{kin}(r_s) = \frac{2.21}{r_s^2} 
\end{equation}

and the correlation energy per particle, in the high density limit, by 
\begin{equation}
\epsilon_{corr}(r_s) = -0.9 + 0.06 \ln(r_s)
\end{equation}

with $r_s$ the Wigner's radius
\begin{equation}
r_s = \left( \frac{3}{4\pi \rho (r)} \right)^{1/3}    
\end{equation}

\begin{figure}[ht!]
    \centering
    \includegraphics[scale=0.6]{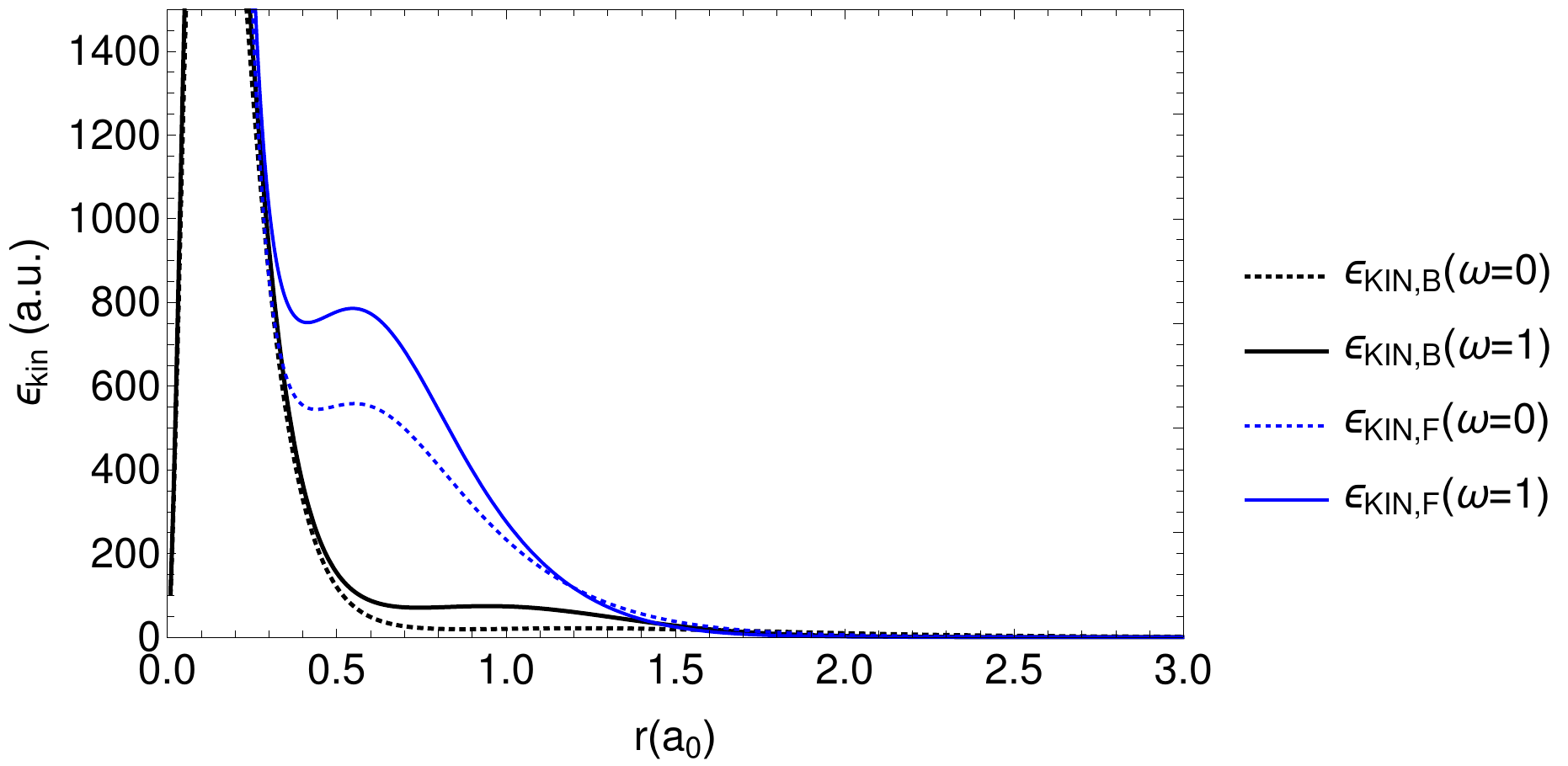}
    \caption{Kinetic energy density of the free and confined Boron and Fluorine atoms.}
    \label{fig:EKinPlot}
\end{figure}

\begin{figure}[ht!]
    \centering
    \includegraphics[scale=0.6]{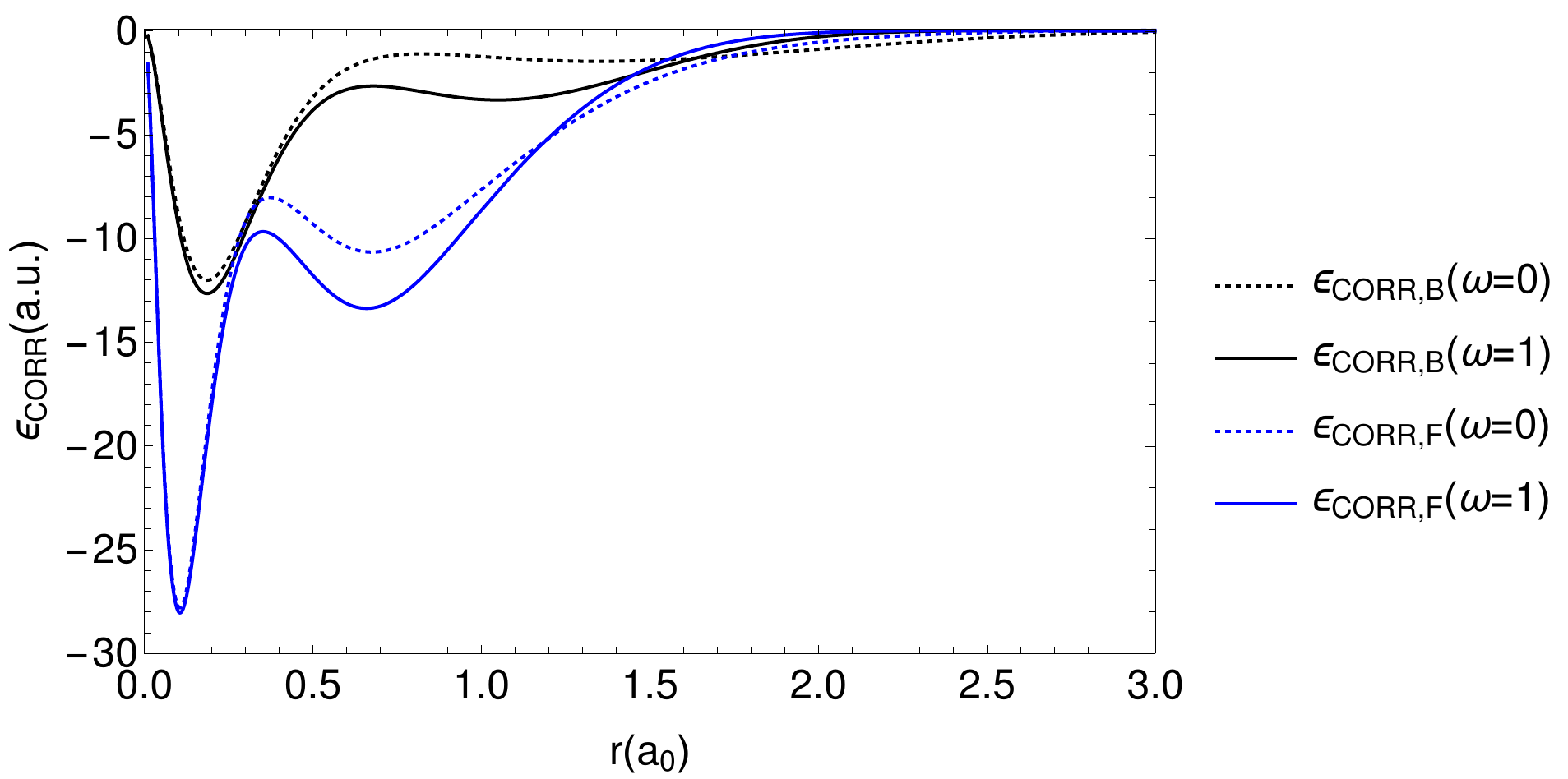}
    \caption{Correlation energy density of the free and confined Boron and Fluorine atoms.}
    \label{fig:ECorrPlot}
\end{figure}

Hence, in the spirit of a local density approximation, the kinetic and correlation energies can be approximated, respectively, by
\begin{equation}\label{eq:Ekin}
E_{kin}= 4\pi\int r^2 \rho (r) \epsilon_{kin} (r_s) dr
\end{equation}

and 
\begin{equation}\label{eq:Ecorr}
E_{corr} = 4\pi \int r^2 \rho (r) \epsilon_{corr} (r_s) dr,    
\end{equation}

where density (or $r_s$) to use in these equations is the one calculated for confined systems. Figures \ref{fig:EKinPlot} and \ref{fig:ECorrPlot} plot the kinetic and correlation energy densities (the argument of the integrals of Eqs. \eqref{eq:Ekin} and \eqref{eq:Ecorr}) as a function of the distance to the nuclei in the B and F atoms. Both cases, the free and the strongly confined atoms ($\omega=1$), are shown. 

One can clearly see the $K$-$L$ shells separation. The correlation energy density is higher in the core region,  which is an expected result. With electrons moving in a small volume,  the average electron-electron distance is small, and the correlation energy is large. Further, in Table 1, the values of kinetic and correlation energies from the valence are also shown. They were obtained integrating Eqs. (6) and (7) only in the valence region demarcated by the ELF function.  Notice that both kinetic and correlation energy, build with the confinement. That is expected within the RPA model because confinement increases the density in every point of the atom.
 
 
However, from closer look to the results, one concludes that the magnitude of the correlation energy proportionally augments more for Boron ($50\%$) than for Fluorine ($11\%$). That reveals in Table 2, where the change in kinetic and correlation energy is shown. Also, the kinetic energy grows faster with confinement in  Fluorine ($\Delta E_{kin}=15.4\ E_h$) than in  Boron   ($\Delta E_{kin}=10.3\ E_h$). These two results imply that in atoms with more electrons, such as O, F, and Ne, electrons move faster, which reduces correlation among them. Contrary, in atoms with fewer electrons, such as Li, B, and C, the increment of kinetic energy is not yet that dramatic and correlation effects keep increasing with the reduction of the average electron-electron distance. Notice that this interpretation clearly explains the changes in the actual correlation energy of Figure \ref{fig:decorr}.
 

\begin{table}[!h]
\centering
\caption{Valence correlation and kinetic energy for B and F atoms (in a.u.)}
\begin{tabular}{cccc}
\hline
  & $\omega$ & E$_{\text{corr}}$ & E$_{\text{kin}}$ \\ \hline
B & $0$      & $-1.93$           & $25.5$           \\
  & $1$      & $-2.90$           & $56.5$           \\ \hline
F & $0$      & $-9.25$           & $363.5$          \\
  & $1$      & $-10.24$          & $471.8$          \\ \hline
\end{tabular}
\end{table}

\section{Conclusions}

In this work, we have studied the effect of confinement, with harmonic potential, on atoms from H to Ne. In many previous works on confinement, electron correlation was neglected or introduced at the LDA and GGA levels of approximation within DFT. Here we performed calculations at the HF, PBE, CCSD, and CASSCF to also understand the importance of electron correlation on the electronic structure of confined atoms. We focused on the following properties: i) changes of electron distribution and electron localization within the $K$ and $L$ atomic shells, ii) confinement-induced ionization pressure, iii) level crossing of electronic states, and iv) the behavior of the electron correlation energy. As for the electron distribution, it is observed that upon confinement, the external $L$ shell is much compressed compared to the internal shell. Also, the position of the maximum of the $L$ shell moves towards the nuclei by non-negligible amounts. However, the size of the inner shell also decreases at strong confinement, and the density at the inter-shells regions increases under pressure. An extreme case is Lithium, for which the separation between the inner and outer shell vanishes at strong confinement $(\omega \ge 0.7 )$. That is revealed by the ELF and the radial distribution function of the electron density (Figure S4). These changes in electron distribution and localization in what is customarily considered the core region of atoms, bring attention to the use of standard pseudopotentials to study materials under extreme pressures. For instance, in Carbon, significant changes in the electron density are observed at distances from the nucleus smaller than the cutoff radius of typical hard PAW pseudopotentials.

A disadvantage of using harmonic confinement is that the strength of confinement cannot be translated into pressure because the volume of the confined atom is not well defined. Here we envisioned a ``calibration'' method, using the ionization pressure of hydrogen, for establishing an adequate volume to convert strength of confinement into pressure. The technique results accurate enough to match the ionization pressure of Lithium with hard-walls confinement. The ionization pressure follows a periodic trend that parallels other properties, such as the ionization potential. The ionization pressure ranges from as low as $28$ GPa for Lithium to $10.8$ TPa for Ne. In HF, the lack of electron correlation tends to underestimate the ionization pressure, while delocalization error in PBE does the opposite.    

In all atoms but Li and Be, slow confinement can be thought of as an adiabatic process in the sense that the nature of the ground state of the neutral atom does not change. However, in Li and Be, low laying excite states, associated with the promotion of $2s$ electrons to $2p$ orbitals, mix with the ground state configuration. There is a critical confinement at which the ground state becomes degenerate. In the case of Be, this happens at $\omega=0.79$, where the singlet configuration $1s^2 2s^2$ becomes degenerate with the triplet $2s^2 2s 2p$. For confinements stronger than $\omega=0.79$, the Hund's rule dictates that the triplet $2s^2 2s 2p$ becomes the ground state. If $\omega$ increases beyond $1.1$, a new crossing occurs and the ground state changes to the triplet $1s^2 2p^2$. In the case of Li, the doublet states $1s^2 2s$ and $1s^2 2p$ become degenerate at $\omega=0.81$. Beyond this point, the ground state is the doublet $1s^2 2p$.  These changes of configuration upon confinement come with potential modifications of the chemical properties of atoms and its molecules. For instance, Be under pressure would be a very reactive species able of $sp$ hybridization that would resemble Boron. Similarly, Li under pressure would readily form antiferromagnetic bonds, which are sustained by the ability of Li to adopt a $1s^2 2p$ configuration.\cite{Shaik2014,cardenas-Li2016}

The effect of confinement on the correlation energy depends on the number of valence electron of the atoms. In the atoms with more electrons, the increment of the kinetic energy with the density becomes dominant over the correlation energy. A semi-quantitative analysis with the RPA model of the kinetic and correlation energy of the HEG (in the high-density limit), shows that in atoms with more electrons, such as O, F, and Ne, electrons move faster, which reduces correlation among them. Contrary, in atoms with fewer electrons, such as Li, B, and C, the increment of kinetic energy is limited, and correlation effects increase with the reduction of electron-electron distance.

\section*{Acknowledgments}
This work was financed by: i) FONDECYT through projects No 1181121 and 1180623, ii) ECOS C17E09, and iii) Centers Of Excellence With Basal-Conicyt Financing, Grant FB0807.

\printbibliography
\end{document}


\maketitle
	
	\listoffigures
	
	\begin{table}[ht]
		\centering
		\label{tb:omega-ion}
		\caption[Ionization confinements for atoms from H to Ne.]{Ionization strengths of confinement for atoms from H to Ne, using different methods of calculation.}
		\begin{tabular}{ccccc}
			\hline
			\multirow{2}{*}{Atom} & \multicolumn{4}{c}{$\omega_\text{ion}$ (a.u.)} \\ \cline{2-5} 
			& HF         & CCSD      & PBE       & Koopmans  \\ \cline{1-5} 
			H                     & $0.80606$  & ---       & ---       & ---       \\
			He                    & $1.39580$  & $1.43653$ & $1.42209$ & $1.43161$ \\
			Li                    & $0.19023$  & $0.19212$ & $0.19918$ & $0.19032$ \\
			Be                    & $0.29431$  & $0.33566$ & $0.32151$ & $0.29917$ \\
			B                     & $0.34172$  & $0.34109$ & $0.35775$ & $0.35623$ \\
			C                     & $0.46197$  & $0.46521$ & $0.47685$ & $0.48487$ \\
			N                     & $0.59878$  & $0.60479$ & $0.61167$ & $0.63048$ \\
			O                     & $0.56638$  & $0.59086$ & $0.61585$ & $0.68682$ \\
			F                     & $0.72513$  & $0.75380$ & $0.76759$ & $0.82535$ \\
			Ne                    & $0.90078$  & $0.93312$ & $0.93714$ & $0.97070$ \\ \hline
		\end{tabular}
	\end{table}
	
	\begin{figure}[ht]
		\centering
		\includegraphics[scale=0.6]{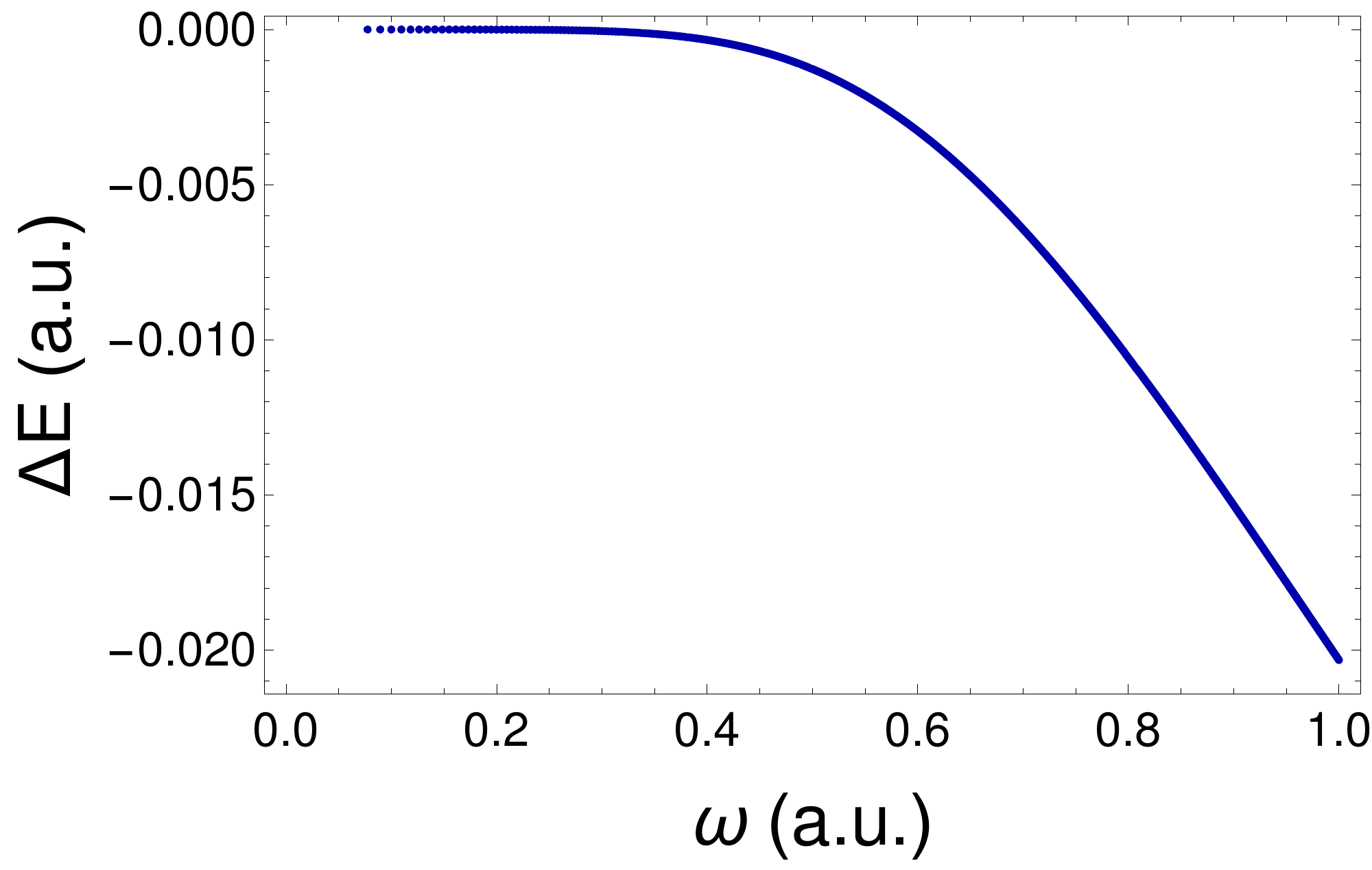}
		\caption[Convergence of the CCSD total energy.]{Convergence of the CCSD total energy with increasing angular momentum of the basis set of the harmonic oscillator.}
		\label{fig:Convergence}
	\end{figure}
	
	\begin{figure}[ht]
		\centering
		\includegraphics[scale=0.8]{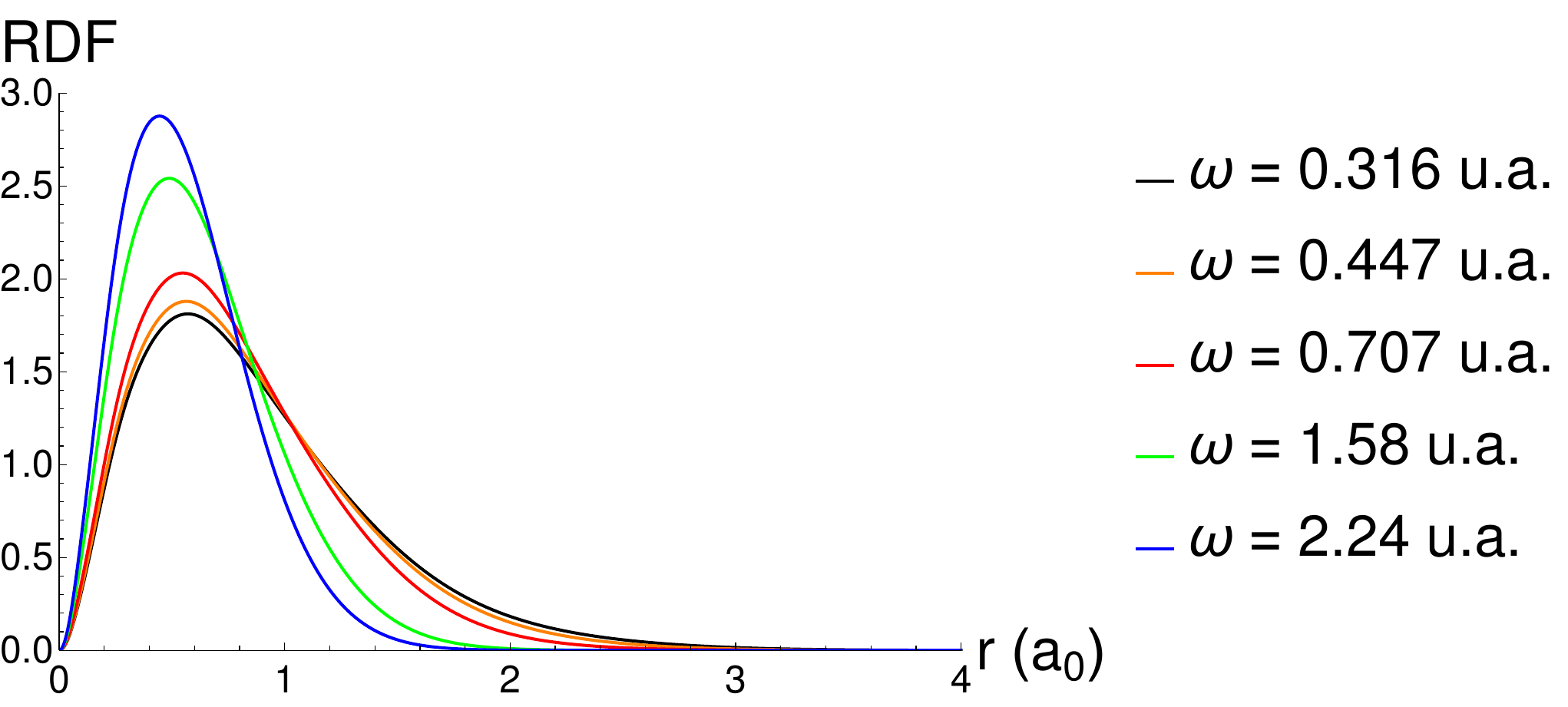}
		\caption[Radial distribution of the confined Helium atom.]{Radial distribution of the Helium atom under different confinements. Red curve is at ionization confinement.}
		\label{fig:RDF_He}
	\end{figure}
	
	\begin{figure}[ht]
		\centering
		\includegraphics[scale=0.8]{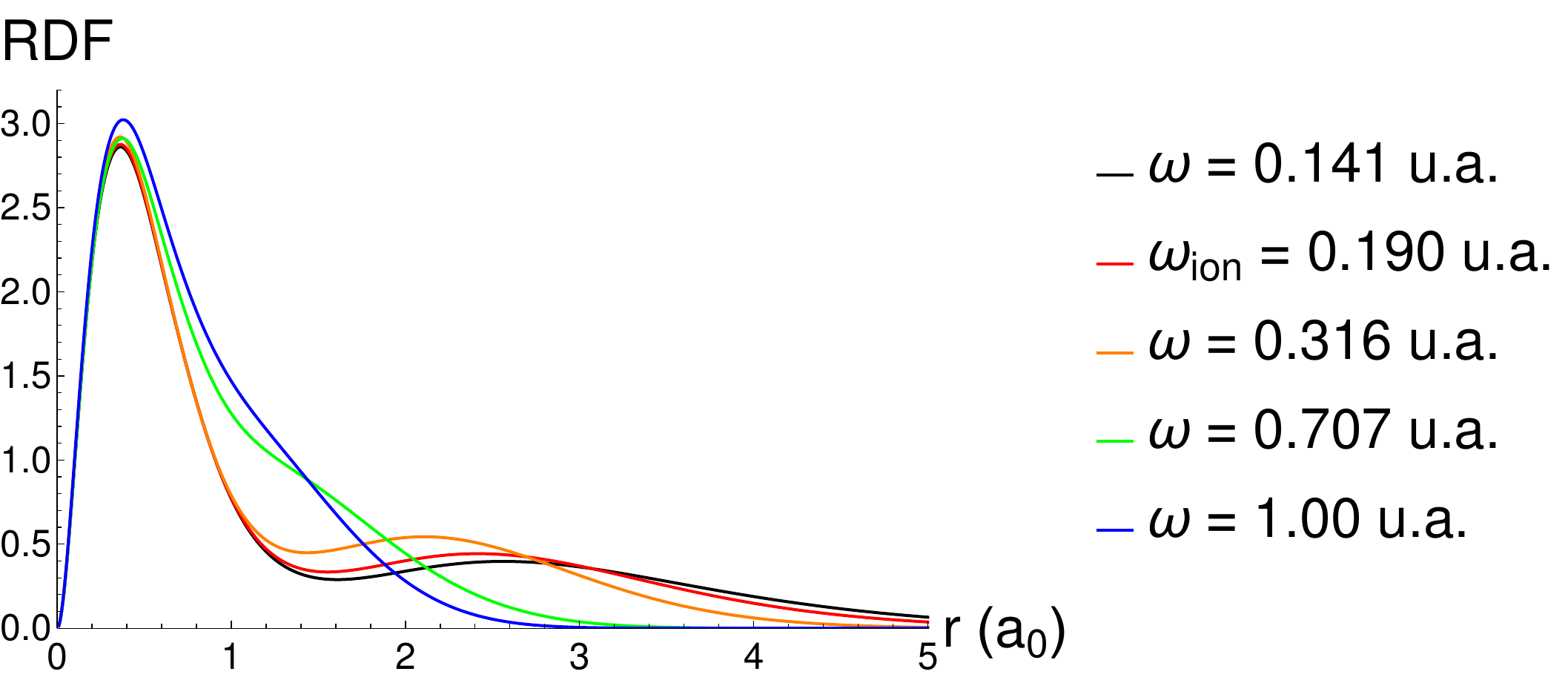}
		\caption[Radial distribution of the confined Lithium atom.]{Radial distribution of the Lithium atom under different confinements. Red curve is at ionization confinement.}
		\label{fig:RDF_Li}
	\end{figure}
	
	\begin{figure}[ht]
		\centering
		\includegraphics[scale=0.8]{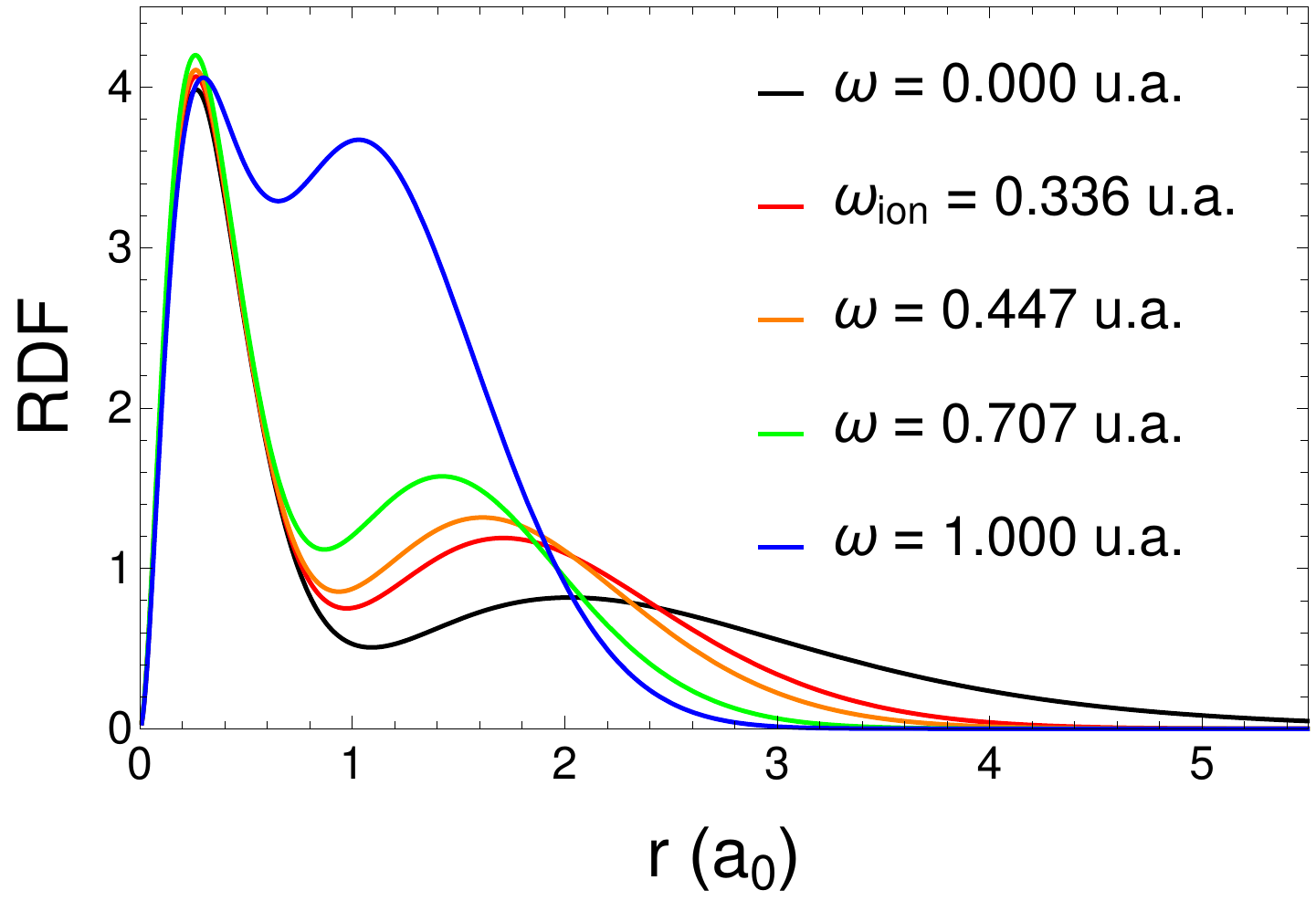}
		\caption[Radial distribution of the confined Beryllium atom.]{Radial distribution of the Beryllium atom under different confinements. Red curve is at ionization confinement. The significant change between the blue curve and the green curve is due to the change in ground state configuration. This occurs at $\omega=0.790$.}
		\label{fig:RDF_Be}
	\end{figure}
	
	\begin{figure}[ht]
		\centering
		\includegraphics[scale=0.8]{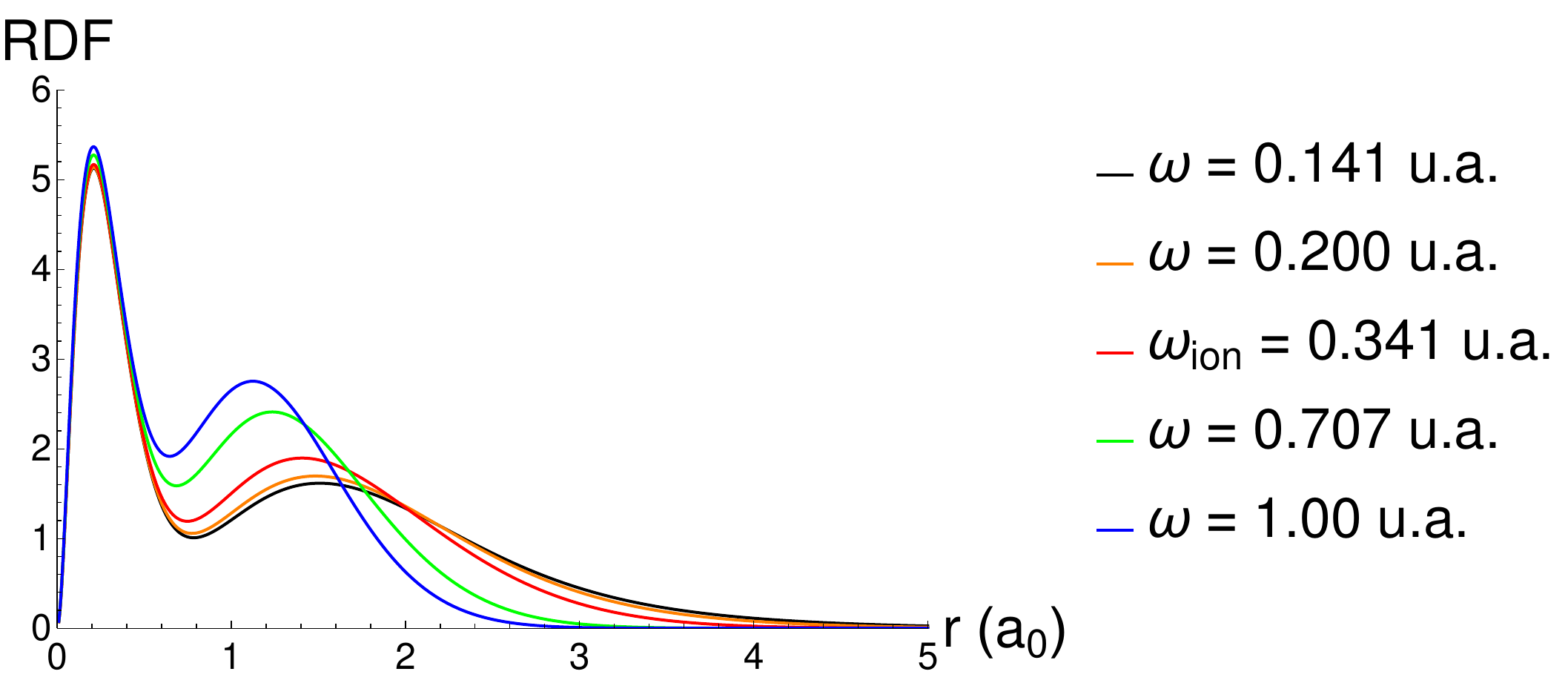}
		\caption[Radial distribution of the confined Boron atom.]{Radial distribution of the Boron atom under different confinements. Red curve is at ionization confinement.}
		\label{fig:RDF_B}
	\end{figure}
	
	\begin{figure}[ht]
		\centering
		\includegraphics[scale=0.8]{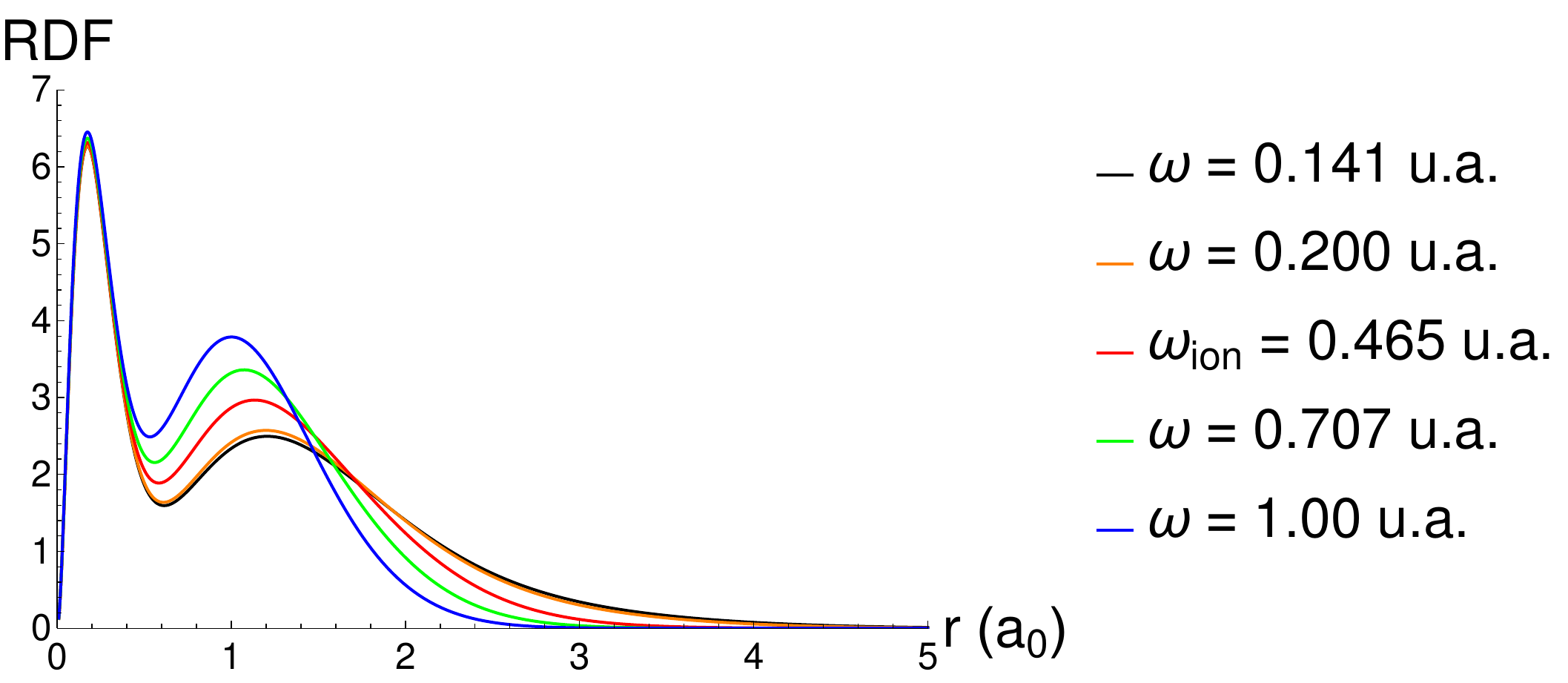}
		\caption[Radial distribution of the confined Carbon atom.]{Radial distribution of the Carbon atom under different confinements. Red curve is at ionization confinement.}
		\label{fig:RDF_C}
	\end{figure}
	
	\begin{figure}[ht]
		\centering
		\includegraphics[scale=0.8]{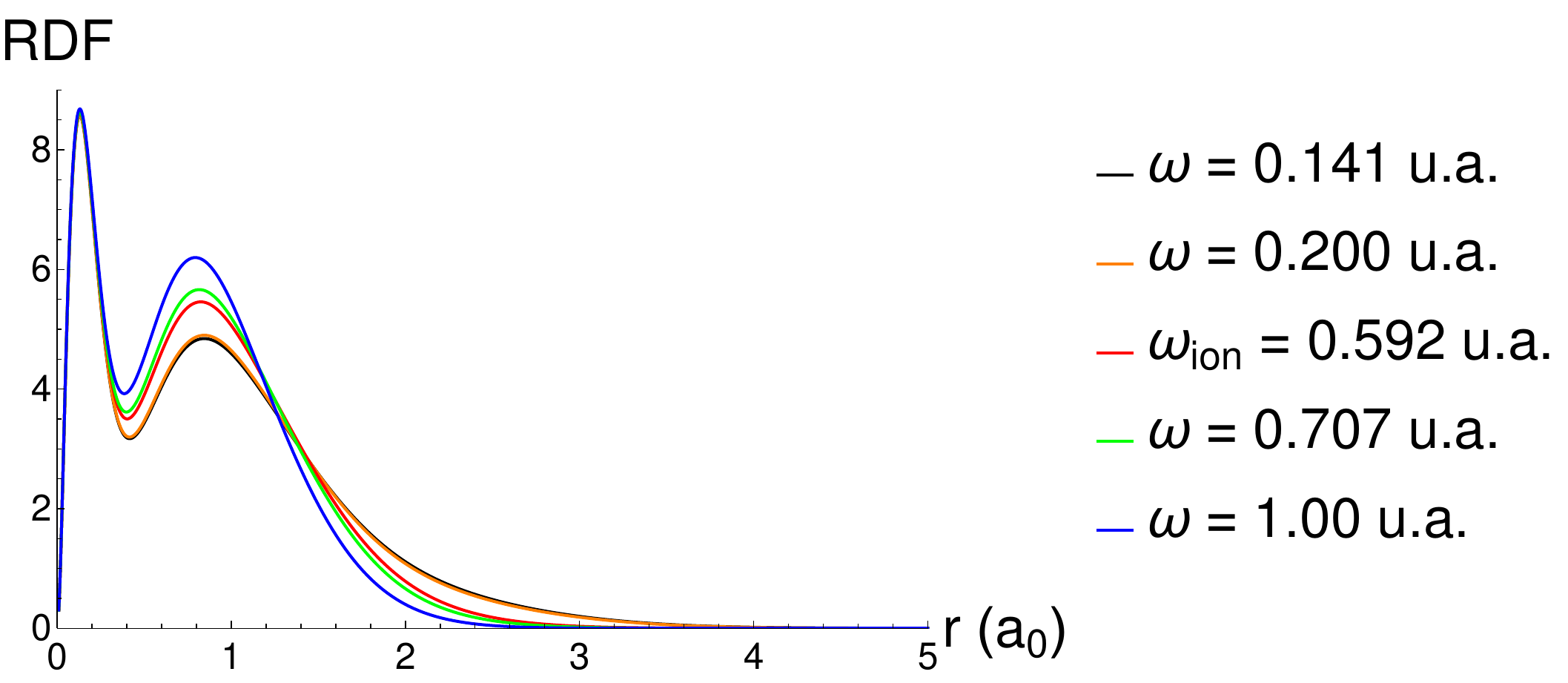}
		\caption[Radial distribution of the confined Oxygen atom.]{Radial distribution of the Oxygen atom under different confinements. Red curve is at ionization confinement.}
		\label{fig:RDF_O}
	\end{figure}
	
	\begin{figure}[ht]
		\centering
		\includegraphics[scale=0.8]{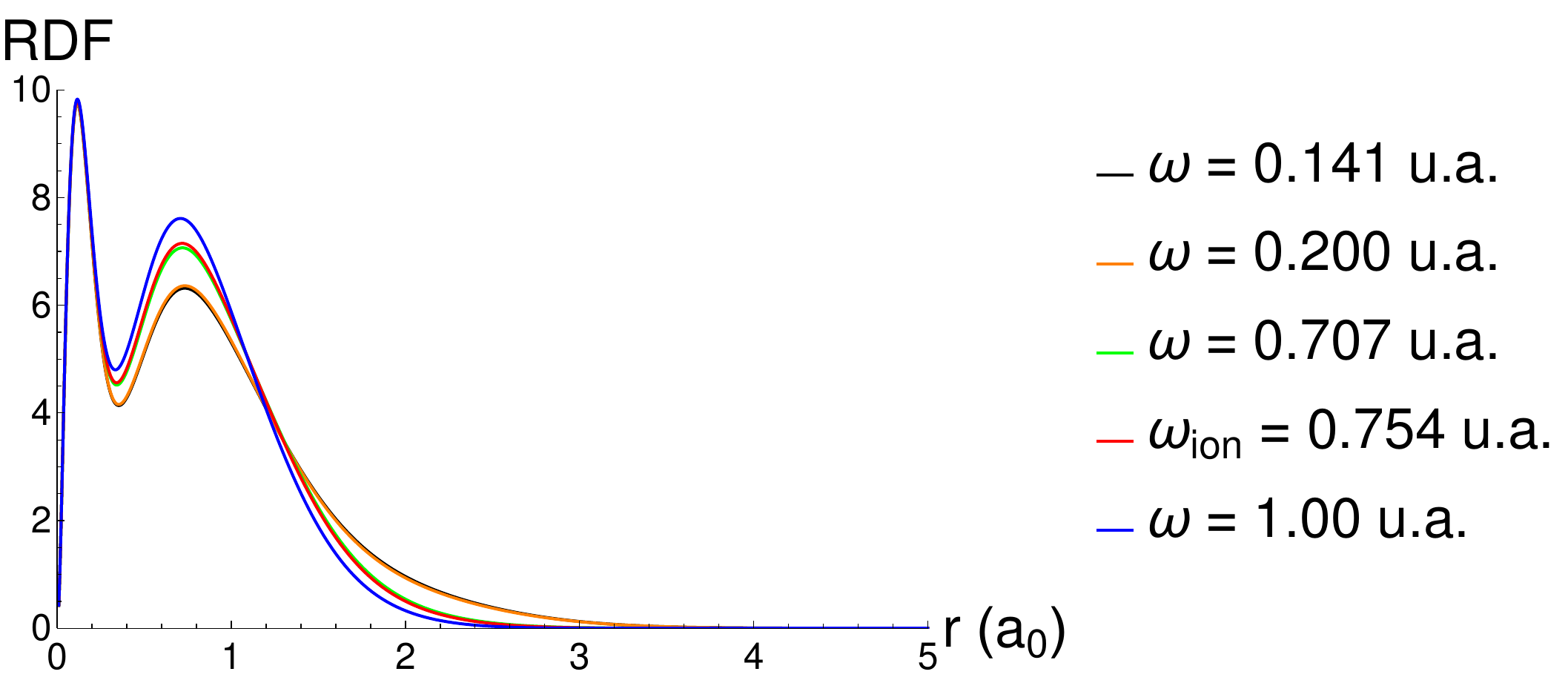}
		\caption[Radial distribution of the confined Fluorine atom.]{Radial distribution of the Fluorine atom under different confinements. Red curve is at ionization confinement.}
		\label{fig:RDF_F}
	\end{figure}
	
	\begin{figure}[ht]
		\centering
		\includegraphics[scale=0.8]{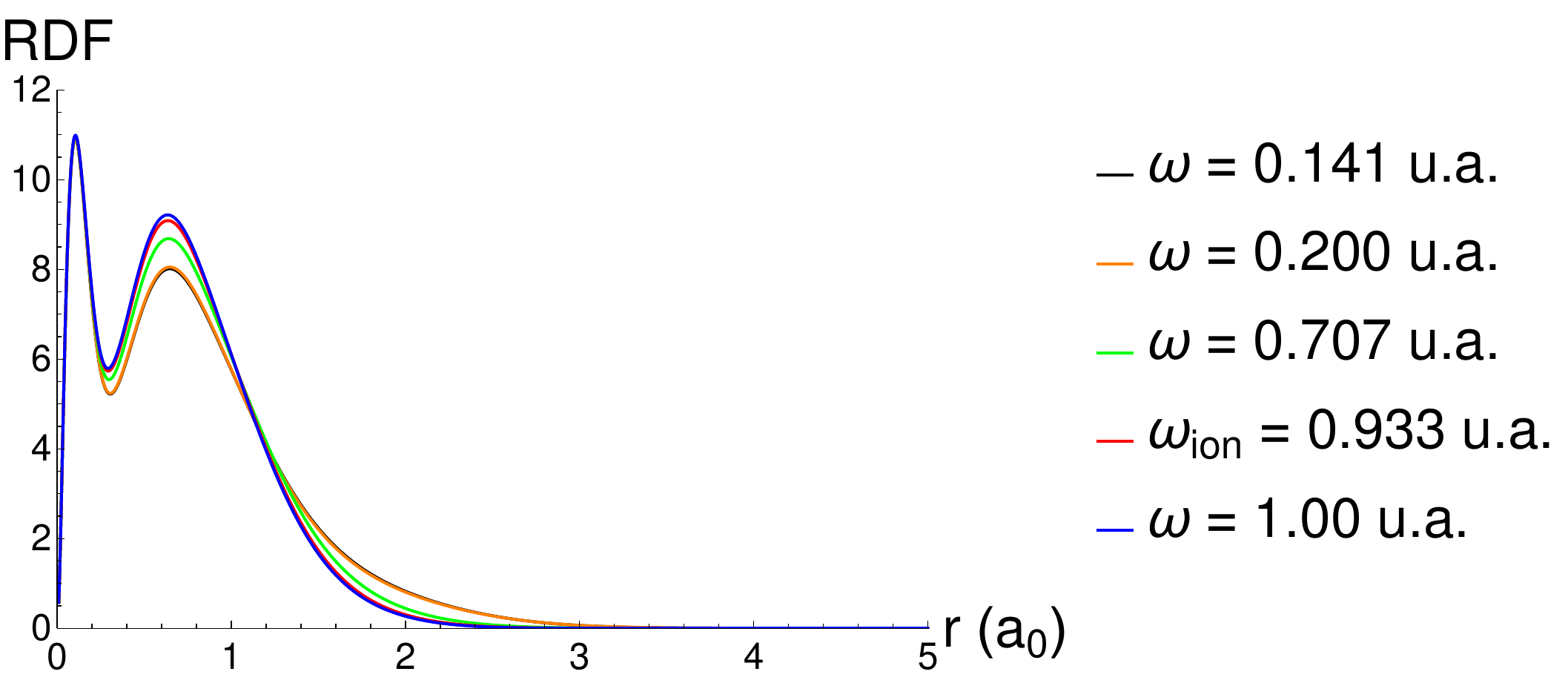}
		\caption[Radial distribution of the confined Neon atom.]{Radial distribution of the Neon atom under different confinements. Red curve is at ionization confinement.}
		\label{fig:RDF_Ne}
	\end{figure}
	
	\begin{figure}[ht]
		\centering
		\includegraphics[scale=0.6]{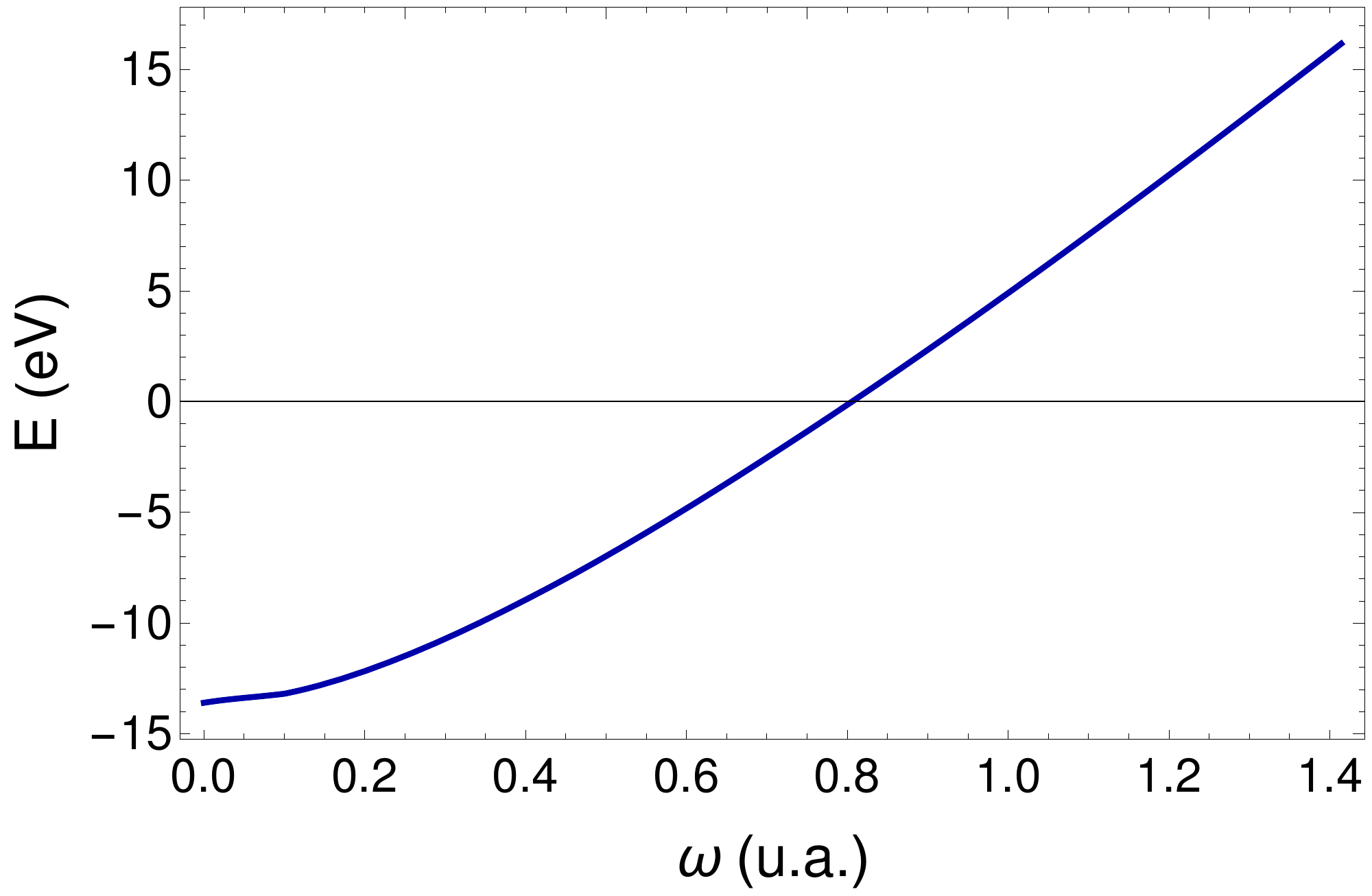}
		\caption{Variation of the Hydrogen's total energy with confinement strength, $\omega$.}
		\label{fig:Ew_H}
	\end{figure}
	
	\begin{figure}[ht]
		\centering
		\includegraphics[scale=1.2]{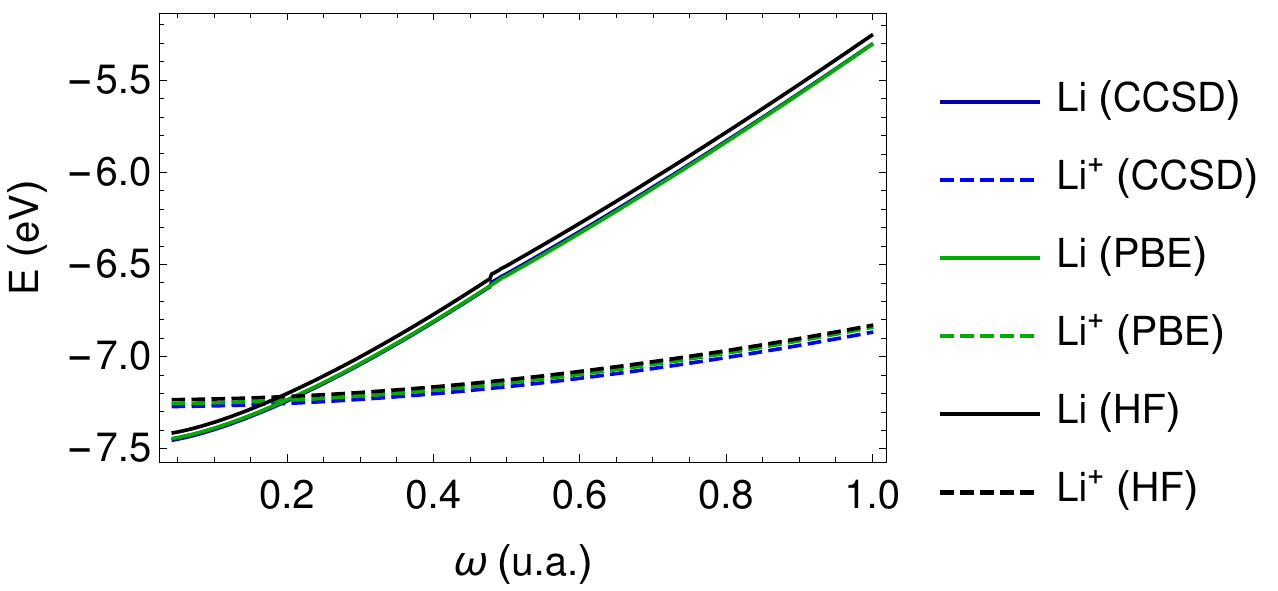}
		\caption[Variation of the Lithium's total energy with confinement strength, $\omega$.]{Variation of the Lithium's total energy with confinement strength, $\omega$. The discontinuities are due to the change in configuration of the ground state of the confined Lithium.}
		\label{fig:Ew_Li}
	\end{figure}
	
	\begin{figure}[ht]
		\centering
		\includegraphics[scale=1.2]{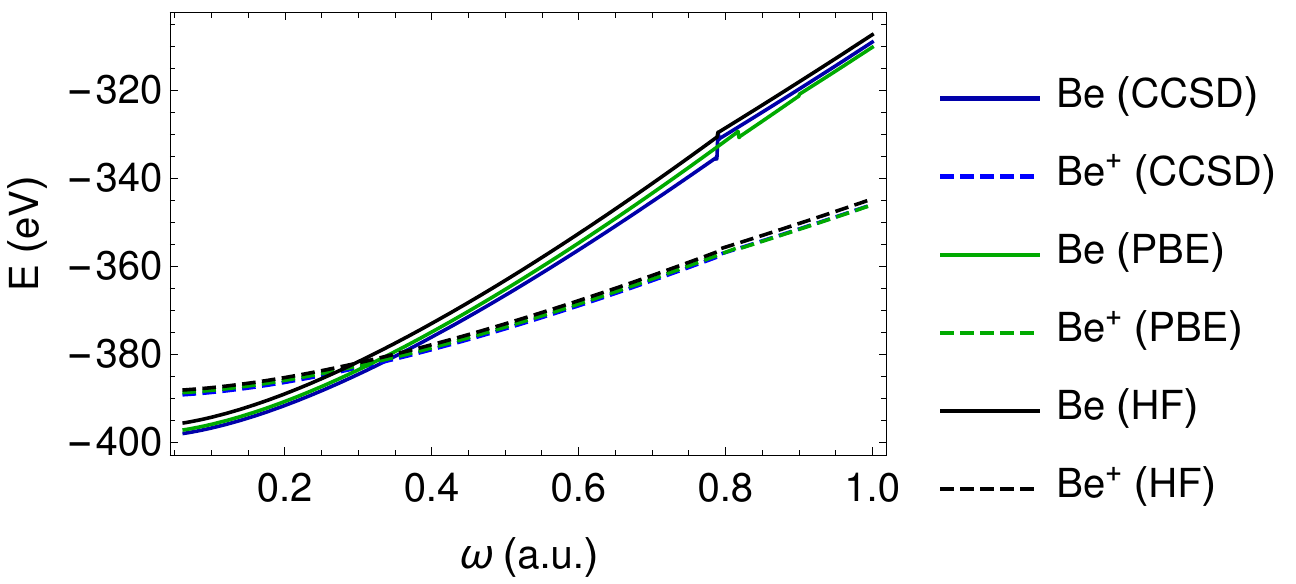}
		\caption[Variation of the Beryllium's total energy with confinement strength, $\omega$.]{Variation of the Beryllium's total energy with confinement strength, $\omega$. The discontinuities are due to the change in configuration of the ground state of the confined Beryllium.}
		\label{fig:Ew_Be}
	\end{figure}
	
	\begin{figure}[ht]
		\centering
		\includegraphics[scale=1.2]{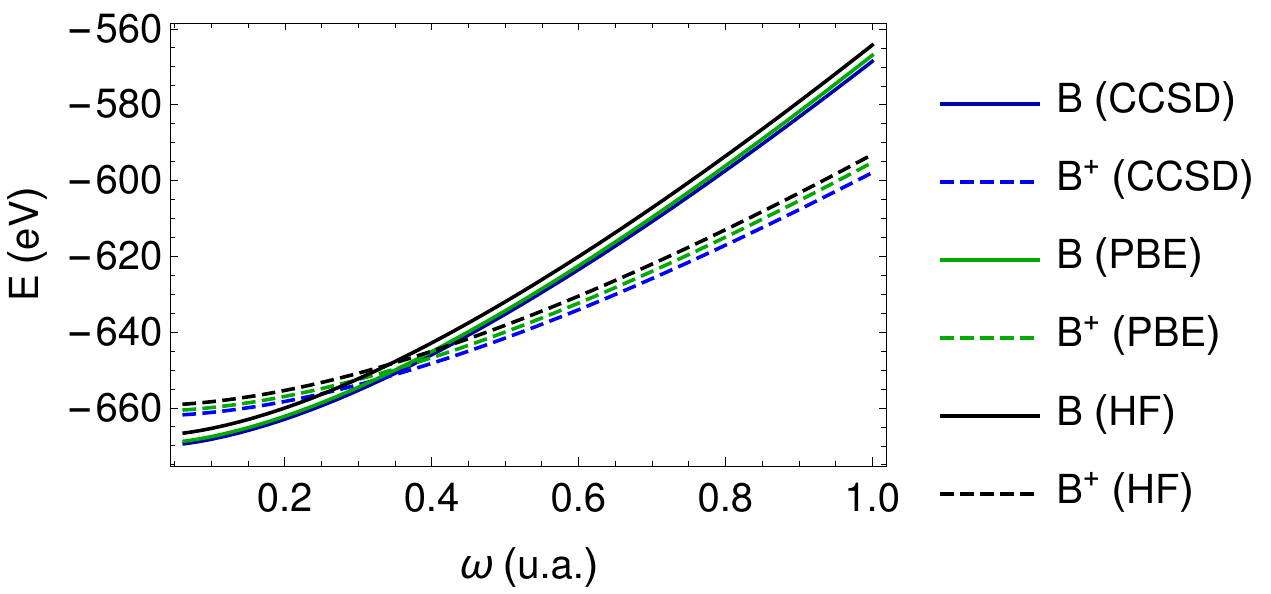}
		\caption{Variation of the Boron's total energy with confinement strength, $\omega$.}
		\label{fig:Ew_B}
	\end{figure}
	
	\begin{figure}[ht]
		\centering
		\includegraphics[scale=1.2]{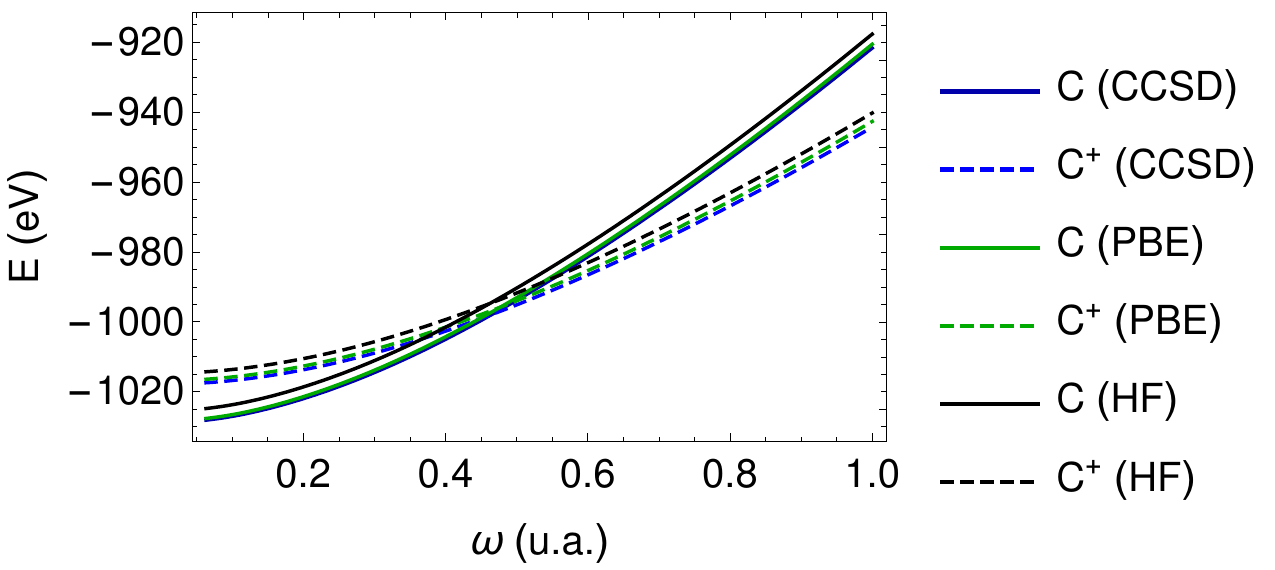}
		\caption{Variation of the Carbon's total energy with confinement strength, $\omega$.}
		\label{fig:Ew_C}
	\end{figure}
	
	\begin{figure}[ht]
		\centering
		\includegraphics[scale=1.2]{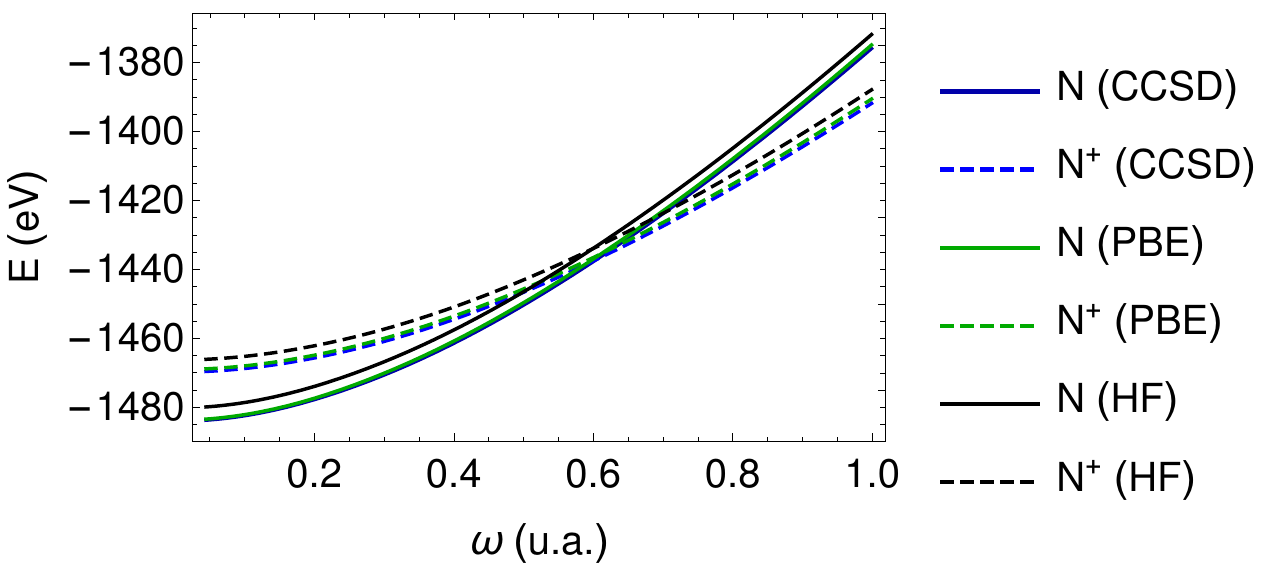}
		\caption{Variation of the Nitrogen's total energy with confinement strength, $\omega$.}
		\label{fig:Ew_N}
	\end{figure}
	
	\begin{figure}[ht]
		\centering
		\includegraphics[scale=1.2]{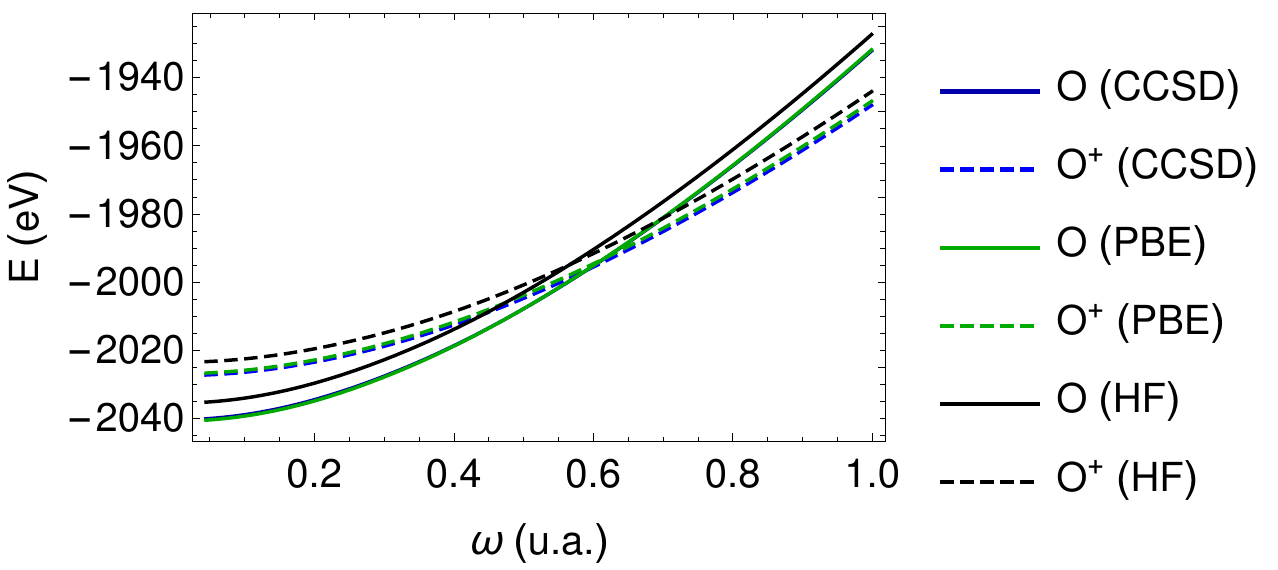}
		\caption{Variation of the Oxygen's total energy with confinement strength, $\omega$.}
		\label{fig:Ew_O}
	\end{figure}
	
	\begin{figure}[ht]
		\centering
		\includegraphics[scale=1.2]{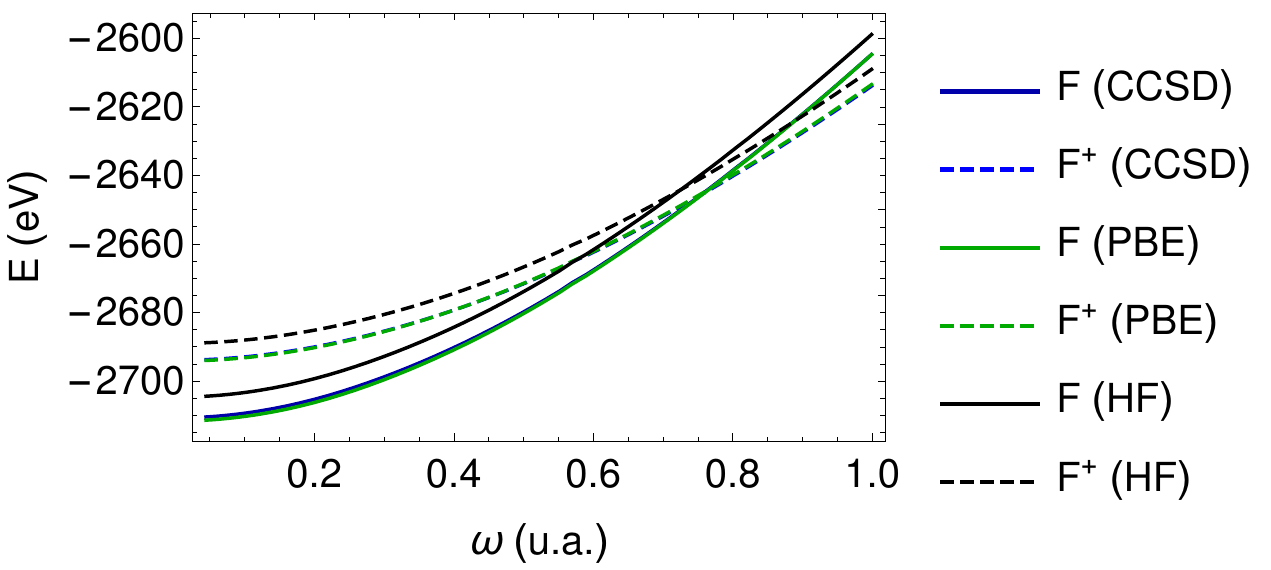}
		\caption{Variation of the Fluorine's total energy with confinement strength, $\omega$.}
		\label{fig:Ew_F}
	\end{figure}
	
	\begin{figure}[ht]
		\centering
		\includegraphics[scale=1.2]{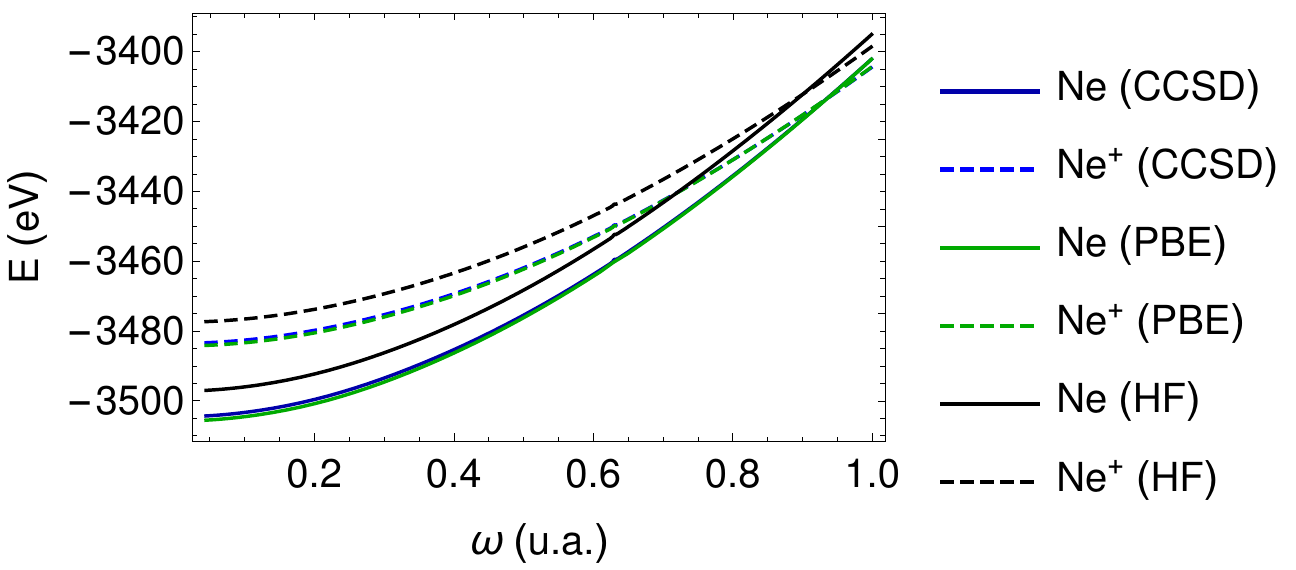}
		\caption{Variation of the Neon's total energy with confinement strength, $\omega$.}
		\label{fig:Ew_Ne}
	\end{figure}
	
	\begin{figure}[ht]
		\centering
		\includegraphics[scale=0.7]{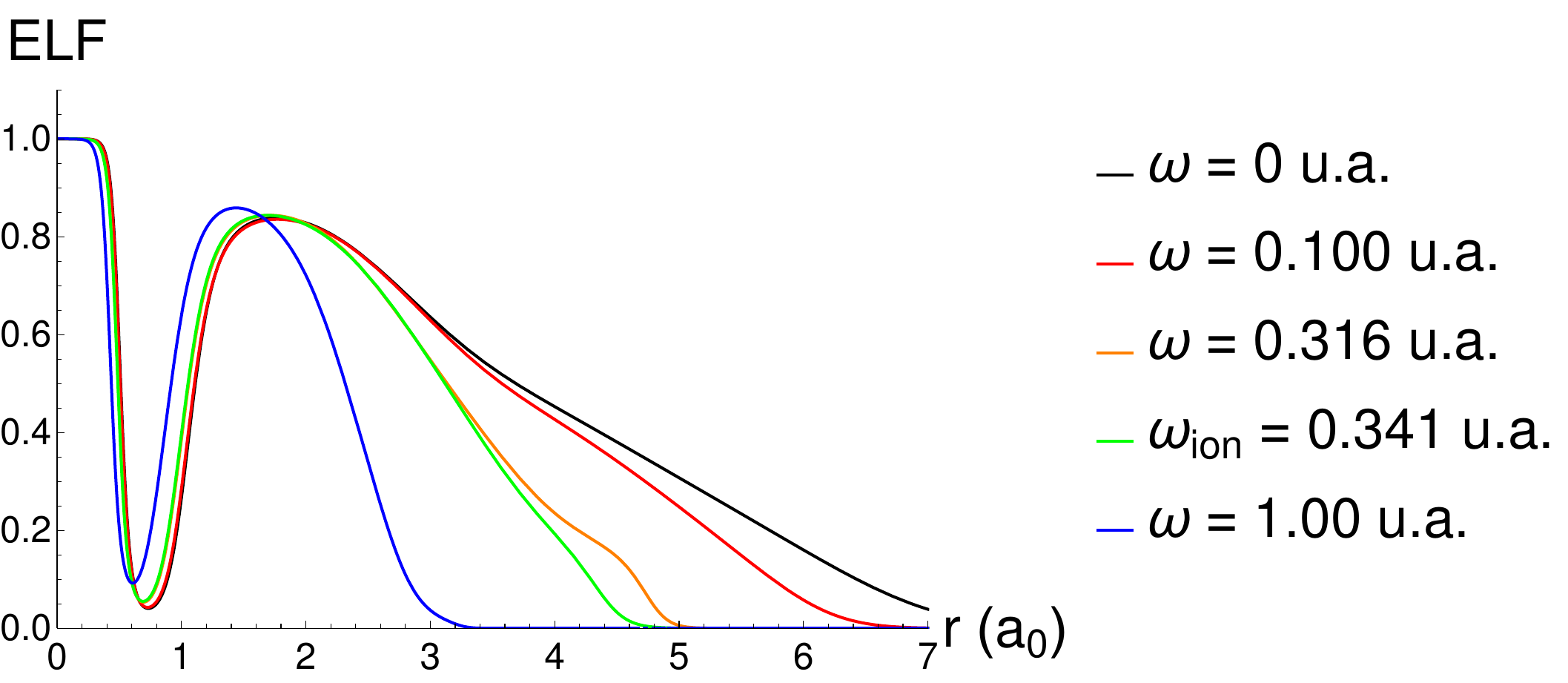}
		\caption{Electron localization function of the confined Boron atom.}
		\label{fig:ELF_B_CCSD}
	\end{figure}
	
	\begin{figure}[ht]
		\centering
		\includegraphics[scale=0.7]{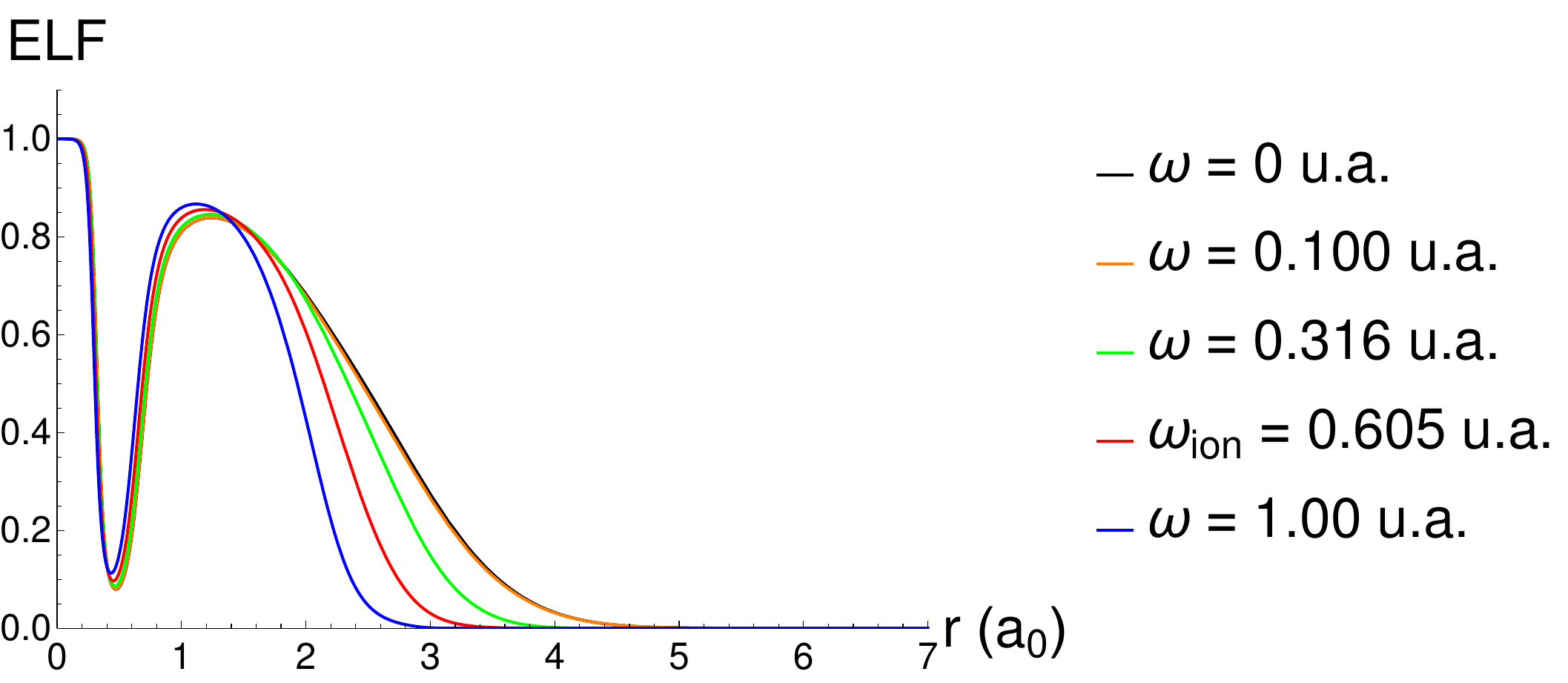}
		\caption{Electron localization function of the confined Nitrogen atom.}
		\label{fig:ELF_N_CCSD}
	\end{figure}
	
	\begin{figure}[ht]
		\centering
		\includegraphics[scale=0.7]{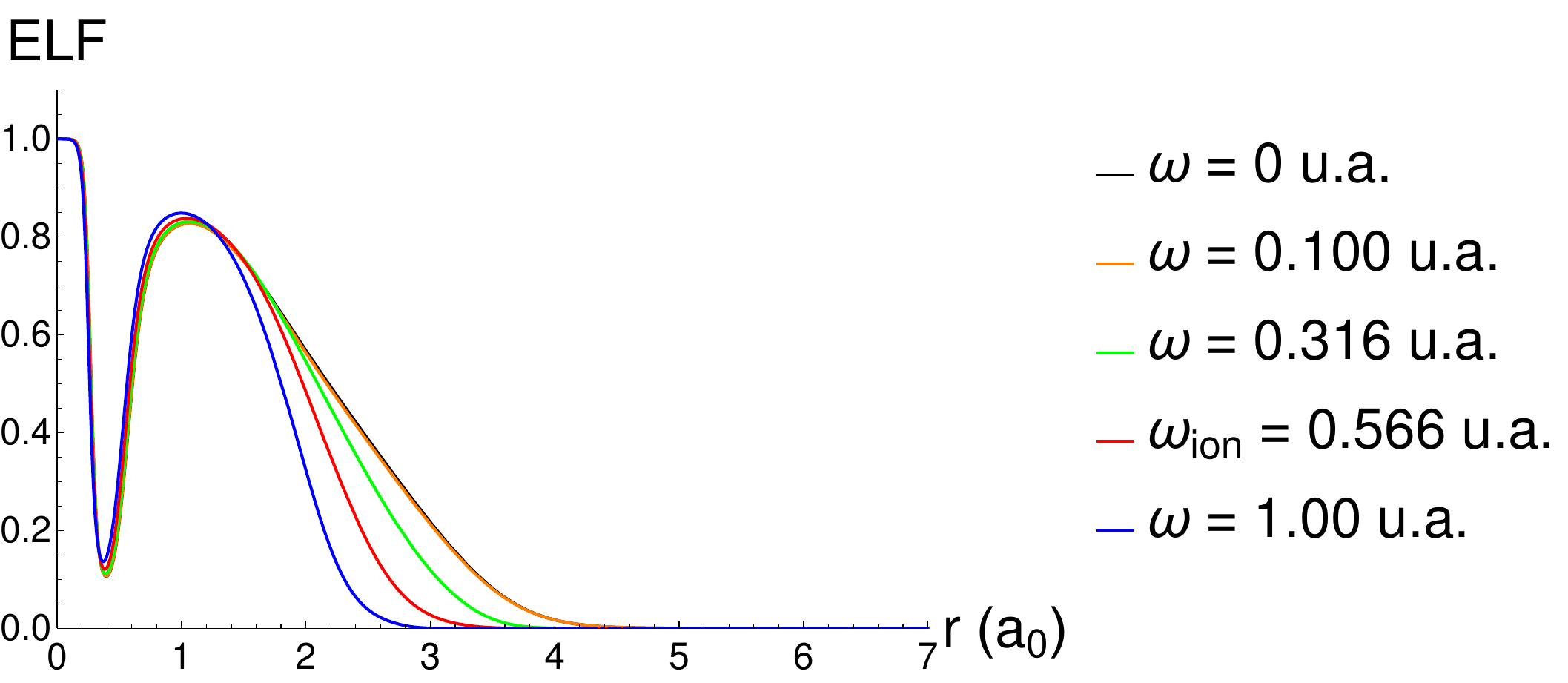}
		\caption{Electron localization function of the confined Oxygen atom.}
		\label{fig:ELF_O_CCSD}
	\end{figure}
	
	\begin{figure}[ht]
		\centering
		\includegraphics[scale=0.7]{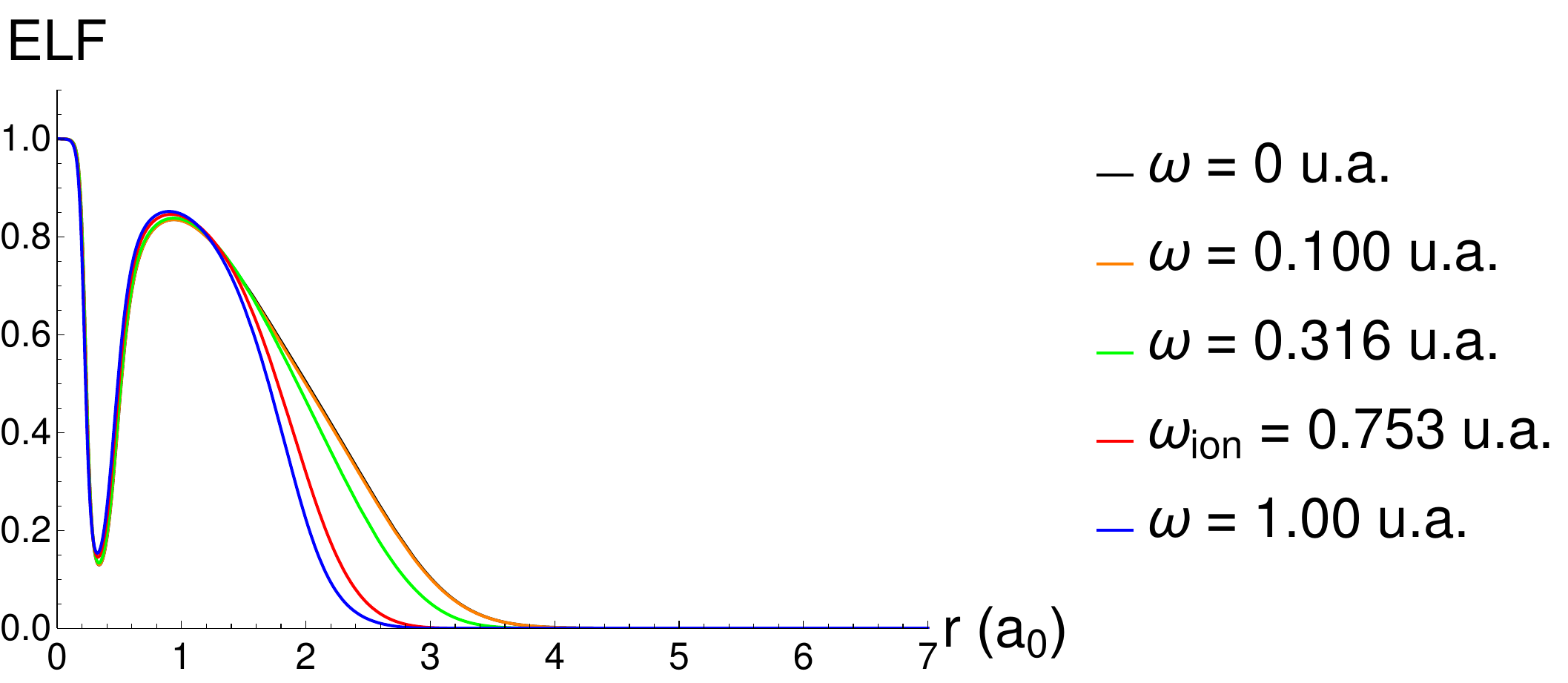}
		\caption{Electron localization function of the confined Fluorine atom.}
		\label{fig:ELF_F_CCSD}
	\end{figure}
	
	\begin{figure}[ht]
		\centering
		\includegraphics[scale=0.7]{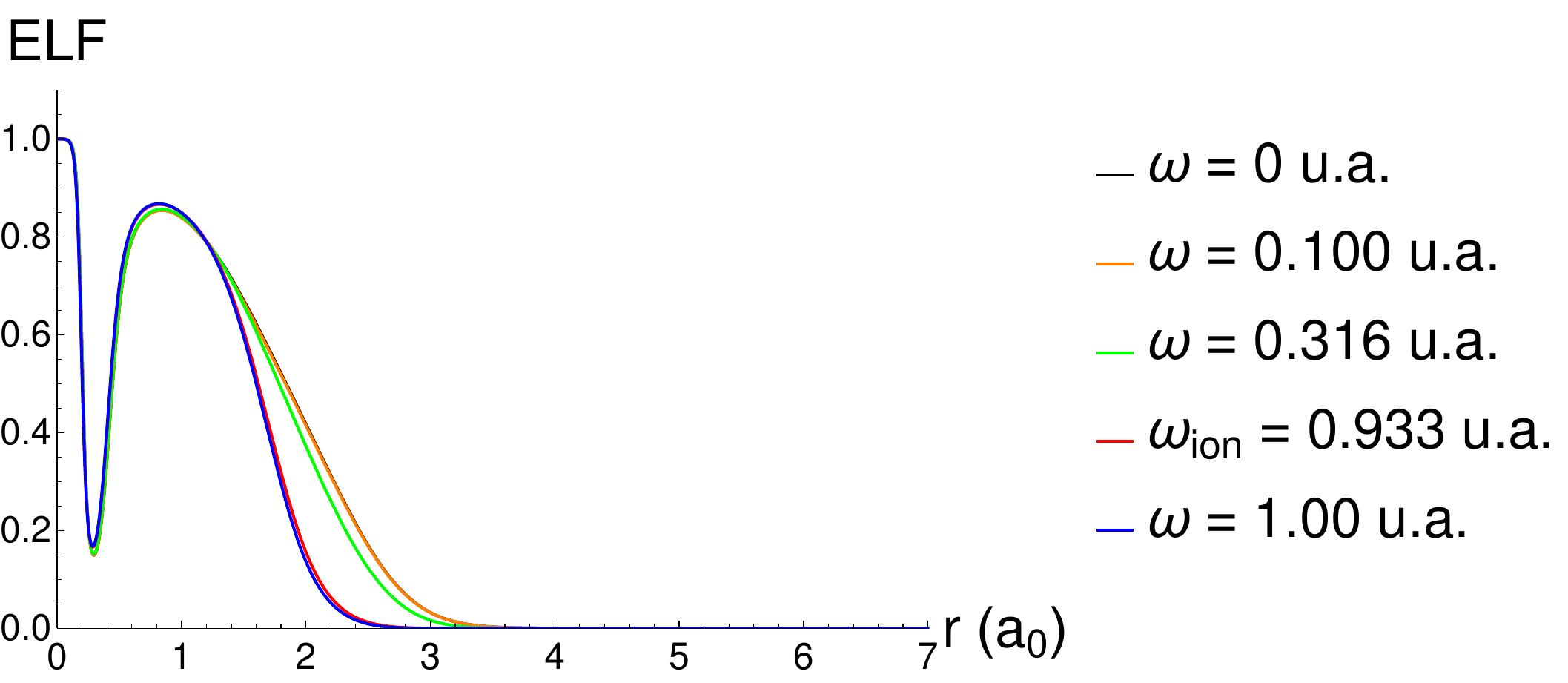}
		\caption{Electron localization function of the confined Neon atom.}
		\label{fig:ELF_Ne_CCSD}
	\end{figure}